\documentclass[12pt,preprint]{aastex}
\slugcomment{Accepted  for publication  in 
 {\it Astronomy \& Astrophysics},  on August 29, 2016}  
\def\lax {\ifmmode{_<\atop^{\sim}}\else{${_<\atop^{\sim}}$}\fi}  
\def\gax {\ifmmode{_>\atop^{\sim}}\else{${_>\atop^{\sim}}$}\fi}  
\def\gtorder{\mathrel{\raise.3ex\hbox{$>$}\mkern-14mu
             \lower0.6ex\hbox{$\sim$}}}

\def\cm2{cm$^{-2}$}
\def\s1{s$^{-1}$}

\begin{document}

\title{ESO 243-49 HLX-1: scaling of X-ray spectral %characteristic 
properties and black hole mass determination}

%\title{4U~1630--47: Spectral state evolution and black hole mass determination}

%\title{Spectral states evolution of 4U~1820-30 observed by BeppoSAX and RXTE: hard X-ray tail (hard X-ray spectral component) detection and stability of spectral index}

%\title{Stability of spectral index of hard comptonized component during state transitions of atoll-source 4U~1820-30 and Z-source GX~340+0}

%\title{Stability of spectral index during slow soft state transitions of atoll-type NS source 4U~1820-30}

%\title{Spectral and timing properties of slow soft state transitions of atoll-type NS source 4U~1820-30}

\author{Lev Titarchuk\altaffilmark{1}  and Elena Seifina\altaffilmark{2} 
%Nikolai Shaposhnikov\altaffilmark{3}
}
%\author{Lev Titarchuk\altaffilmark{1}, Elena Seifina\altaffilmark{2} and Filippo Frontera\altaffilmark{3}  }
\altaffiltext{1}{Dipartimento di Fisica, Universit\`a di Ferrara, Via Saragat 1, I-44122 Ferrara, Italy, email:titarchuk@fe.infn.it}
\altaffiltext{2}{Moscow M.V.~Lomonosov State University/Sternberg Astronomical Institute, Universitetsky 
Prospect 13, Moscow, 119992, Russia; seif@sai.msu.ru}
%\altaffiltext{3}{CRESST/University of Maryland, Department of Astronomy, College Park MD 20742, 
%Goddard Space Flight Center, NASA,  code 663, Greenbelt  
%MD 20771,  USA: email:nikolai.v.shaposhnikov@nasa.gov USA
%}
%\altaffiltext{2}{Dipartimento di Fisica, Universit\`a di Ferrara, Via Saragat 1, I-44122  Ferrara, Italy, email:frontera@fe.infn.it

%}
\begin{abstract}
{We report the results of $Swift$/XRT  observations (2008$-$2015) of a  {hyper}-luminous X-ray source, ESO 243-49 HLX-1. 
We found %show   
a strong observational evidence that  ESO 243-49 HLX-1 underwent  %undergoes  
spectral transitions from the low/hard  state 
to the high/soft state during these observations. The spectra of ESO 243-49 HLX-1 are well fitted by the so-{called} bulk motion Comptonization  model  for all spectral states. 
%We have also investigated a possible absorption effect in X-ray spectra of M101 ULX-1 
%(from neutral matter) taking into account two possible cases for the absorbing column $N_H$: 
%($i$) the constant  %along state transition 
%and ($ii$) the variable of $N_H$ %ranging from 0.1 to $5\times 10^{21}$ cm$^{-2}$ 
%when the source undergoes  the  state transition. 
We have established the photon index ($\Gamma$) saturation level,
$\Gamma_{sat}$=3.0$\pm$0.1, in the correlation of $\Gamma$ versus  mass accretion rate ($\dot M$). 
This $\Gamma-\dot M$ correlation allows us to estimate  the black hole (BH) mass in ESO 243-49 HLX-1 to be  
$M_{BH}\sim 7\times 10^4 M_{\odot}$, assuming the %spread in 
distance  to ESO 243-49 of 95 Mpc. %(from $95.4\pm 0.5$ Mpc to $??.?\pm 0.6$ Mpc). 
For the BH mass estimate we used the scaling method, taking Galactic BHs  XTE~J1550-564, 
H~1743-322 and 4U~1630-472, and an extragalactic BH source, M101 ULX-1 as reference sources. The $\Gamma-\dot M$  
%$\Gamma$ versus $\dot M$ 
correlation revealed in ESO 243-49 HLX-1 is similar to those in a number of Galactic and extragalactic BHs and 
it  clearly shows the correlation along with 
the strong $\Gamma$ saturation at  $\approx 3$. This is  a  {reliable} %observational  
observational evidence of a BH in ESO 243-49 HLX-1. We also found that  the  seed (disk) photon 
temperatures are quite low, of order of 50$-$140 eV which are consistent with  a high BH mass in ESO 243-49 HLX-1. 
%Our estimate of the BH mass in ESO~243-49 HLX-1 is consistent with the recent IMBH mass estimate using the 
%the soft-to-hard state transition luminosity of HLX-1
%luminosity measurements of this IMBH candidate by Yan et al. (2015), Servillat et al. 2011, Farrell et al. 2009 and % and Webb et al. (2012 ???), 
%Soria et al. (???), who find that $M_{BH}=(8\pm 4) \times 10^4$ solar masses.
%This is consistent with the IMBH mass of ?104-105MS derived from detailed X-ray
%spectral modelling (Farrell et al. 2010; Davis et al. 2011;
%Servillat et al. 2011; Godet et al. 2012; Webb et al. 2012
%We obtain that the central object in ESO 243-49 HLX-1 has intermediate BH mass of order %of 10$^{5}$ solar masses.
}

\end{abstract}


\section{Introduction}
The prominent edge-on galaxy ESO 243-49, located at 95 Mpc  away in the constellation Phoenix (see Afonso et al. 2005),  { hosts a hyper-luminous X-ray source commonly known as HLX-1, which is possibly an intermediate-mass black hole (IMBH).}
This black hole candidate (BHC)  was first observed in 2004 as a source emitting X-rays in the vicinity of the spiral galaxy 
ESO 243-49 and it was catalogued as 2XMM J011028.1-460421.  Later, in 2008 the field of this X-ray source was deeply re-imaged 
by a team  led by Natalie Webb (the Institut de Recherche en Astrophysique et Planetologie in Toulouse, France).

Farrell et al. 2009  suggested that HLX-1 is an intermediate-mass black hole candidate with  mass of $\sim 10^4$ M$_{\odot}$  because of  very high X-ray luminosity of HLX-1 ($\sim 10^{42}$ erg/s in the 0.2 -- 10 keV range)  and because  
its disk blackbody spectrum peaked at $kT_{in}\sim$2 keV along with its X-ray spectral  variability.  X-ray spectral analysis (see Godet et al. 2009, 2012; 
Servillat et al. 2011; Lasota et al. 2011; Farrell et al. 2009; Davis et al. 2011), optical observations  (see Farrell et al. 2012; Wiersema et al. 2010; 
Soria et al. 2010) and radio observations (see Webb et al. 2012)  support the  presence of  an intermediate-mass BH   in ESO 243-49  HLX-1. 

Long-term monitoring with the $Swift$/XRT has shown that X-ray luminosity of HLX-1 changes by a factor of $\sim$50
(Godet et al. 2009; Yan et al. 2015), and its %correlated 
spectral variability is reminiscent of that seen in Galactic stellar-mass BHs
(see Servillat et al. 2011). %Since 
Specifically, the X-ray light curve demonstrates a recurrent fast rise-exponential decay'' (FRED) type of pattern  
%which  a quasi-regular periodicity 
 in the range of $\sim$370 -- 460 days. This  recurrency has been interpreted as an
orbital period of the companion star (see Lasota et al. 2011 and Soria 2013). 
However, the last 2 outbursts have been  too late to be
consistent with that explanation. The interval between the last two outbursts is
almost 3 months longer than the interval between the first two [Soria (2015),  a private communication].

Recently, Webb et al. (2012), hereafter W12, reported a detection of transient radio emission at the location of HLX-1, which  agrees  with a discrete jet ejection event. These observations also allowed W12 to re-estimate  the BH mass   to be between $\sim$9$\times$10$^3$ M$_{\odot }$ and 
$\sim$9$\times$10$^4$ M$_{\odot }$. %However, these radio observations %of Galactic globular clusters 
%currently provide only upper limits on the mass. % (e.g. Maccarone & Servillat 2008; Lu & Kong 2011). 
In contrast, King \& Lasota (2014) suggested that HLX-1 may be a stellar mass binary like
SS~433 (see also Lasota et al. 2015), in which the X-ray emission possibly comes from the beamed jet. 

HLX-1 is projected in the sky at $\sim$0.8 kpc out of the plane and $\sim$3.3 kpc ($\approx$8 {\tt "})
%arcseconds) % north-east) 
of the nucleus of the S0/a
galaxy ESO 243-49 (the luminosity distance $\sim$ 96 Mpc). 
Galaxy ESO 243-49 is a member of the cluster Abell 2877
(e.g. Malumuth et al. 1992). The association of HLX-1 with
ESO 243-49 is confirmed by the redshift measurements of the observed
H$_{\alpha}$ emission line of the counterpart (Soria et al. 2013, hereafter SHP13; Wiersema et al. 2010),
although the velocity offset between this and the bulge of ESO 243-49 is $\sim$430 km/s, close to the escape velocity
from the S0 galaxy (SHP13). This  allows to suggest that HLX-1 might be in a dwarf satellite galaxy  or a star cluster near ESO 243-49 are not than in the galaxy itself (SHP13). The HII region in which the HLX is located could be the remnant of a dwarf satellite galaxy that has been accreted (Farrell et al. 2012). The optical counterpart of HLX-1 was detected in various bands, from near-infrared to far-ultraviolet (FUV, Wiersema et al. 2010; Soria et al. 2010, 2012; Farrell et al. 2012), but its nature remains puzzling. 

 Cseh et al. (2015) used radio observations of ESO 243-49 HLX-1 in 2012
based on the Australia Telescope Compact Array (ATCA) and Karl G. Jansky Very Large Array
(VLA). 
%with an assumption of the called ``Fundamental Plane’' approach 
They esimated the BH mass as
$2.8^{+7.5}_{2.1}\times 10^6$ M$_{\odot}$. Yan et al. (2015) have analyzed $Swift$ monitoring observations of ESO 243-49 HLX-1 and  compared the HLX-1 outburst properties with those of bright Galactic low-mass X-ray binary transients (LMXBTs). 
Furthermore, they  stated that HLX-1 spends progressively less time in the succeeding outburst state and  much more time in quiescence, but its peak luminosity remains approximately constant. The spectral analysis by Yan et al.  strengthened the similarity between the state 
transitions in HLX-1 and those in Galactic LMXBTs. 

A very high luminosity of ESO~243-49 HLX-1 is considered as  
evidence for the existence of IMBH in HLX-1 (Farrell et al. 2009). However, luminosities 
up to $\sim$10$^{41}$ erg/s can be explained by stellar-mass BHs undergoing super-Eddington accretion (see Begelman 2002) that is followed.
 As a result,  the apparent luminosity  
%which makes them 
can  exceed the Eddington limit estimated 
for isotropic radiation (King 2008; Freeland et al. 2006). 
Highly collimated sources are not expected to have a disk-blackbody or thermal-dominant spectrum, like we see in HLX-1 in the high/soft state. 
Therefore HLX-1 may not be a strongly beamed source.

Consequently, luminosity above $\sim 10^{41}$ erg/s
is difficult to explain without considering   a massive BH source. Generally, two scenarios for an interpretation of HLX/ULX phenomena  have been proposed. 
%and its high luminosities over of $10^{39}$ erg/s it is presently contemplated two proposed scenarios for ULXs. 
 First, these sources 
%ULXs have been associated with supernovae, many are thought to be accreting objects. More specifically, 
%they 
can be stellar-mass BHs ($<$100 M$_{\odot}$) %significantly less than 100 solar masses ($M_{\odot}$)] 
radiating at Eddington  or super-Eddington rates [%\cite{tl97}, 
see e.g. \cite{Mukai05}]. 
Alternatively, they can be intermediate-mass BHs ($>$100 M$_{\odot}$) %(more than 100 $M_{\odot}$) 
whose  luminosities are essentially 
%radiating 
sub-Eddington.
% regimes. 
The exact  origin of such objects still remains uncertain, and there is still no general consensus on the causes of the  poweful outbursts. 
%{\bf
Recently, \cite{Bachetti14} discussed another scenario
%/nature 
for ULX. 

Using the {\it NuSTAR} mission { \cite{Harrison13}}, Bachetti
and collaborators  detected pulsations of X-ray emission with an average period 1.37 s  modulated by a 2.5-day 
%sine 
cycle % period % modulation from %bright 
from ULX-4 located in the nuclear region of the galaxy M82. 
%using $Nu$STAR mission~\citep{Harrison13}. 
Bachetti et al.  also argued that these pulsations are results of  
the rotation of a magnetized neutron star, while the modulation arises from its binary orbital motion.  
We note that the X-ray luminosity of M82 ULX-4 is about $L_X \sim 2\times 10^{40}$ erg s$^{-1}$ in 0.3 -- 10 keV 
energy range, which suggests a luminosity $\sim 100\times L_{Edd}$ %times the Eddington limit 
for a 1.4 M$_{\odot}$ neutron star (NS). Such a  source  is   ten times  brighter than any  known 
accreting pulsar.  The discovery of M82 ULX-4   and  its possible interpretation as a NS  
 %that neutron stars are also may related to %not be rare in the 
%ULX population, which 
can expand   possible scenarios for ULXs.
%and it challenges physical models for the accretion of matter 
%onto magnetized compact objects.      
%}
 
%None of those scenarios 
%appears fully consistent with all the observational constraints, and there is still no general consensus on
%what causes the outbursts. 

It is  desirable for ESO~243-49 HLX-1 to independently identify the BH for its central  
object  
%it is  desirable to have an independent BH identification for its central  
%object  
%located  in the center of ESO~243-49 HLX-1 
and also determine the mass of its   BH as an alternative to previously employed methods (based on luminosity estimates alone).
A new method for determining  the BH mass 
%of  the BH mass determination 
was developed by Shaposhnikov \& Titarchuk (2009), hereafter ST09, who used a correlation scaling between X-ray spectral and timing (or mass accretion rate) properties observed from many Galactic BH binaries during 
the spectral  state transitions. 
%{\bf
%Application of this method has  been also extended 
This method has   also been applied  to  study of another class of X-ray sources, ULXs sources,  NGC 5408 X-1 (Strohmayer \& Mushotzky (2009) and M101 ULX-1 
(Titarchuk \& Seifina, 2015). We note, this method is commonly used for a BH mass determination of supermassive BHs, 
such as %Mrk 766, 
NGC 3227, NGC 5548 NGC 5506 and NGC 3516 (Papadakis et al. 2009; Sobolewska \& Papadakis, 2009) and  NLS1 galaxy 
{Mrk 766} (Giacche et al. 2014), 
%, A&A, 494, 905 [Mrk 766, NGC 3227, NGC 5548 NGC 5506 and NGC 3516], Giacche et  al., A&A, 2014, 562, 44 [NLS1 galaxy Mrk 76], 
using a sample of Galactic BHC binaries as reference sources.  %A good agreement of their results for sources with that determined by alternative methods 
%known values of BH masses and distance 
%provides independent verification for our scaling technique. 
%}
This scaling  method   can also  be applied for  BH mass  estimates  
%of the central object (for e xample, a BH mass) based on X-ray data even
 when  the conventional dynamical method cannot be used.

We applied the ST09 method to {\it Swift}/XRT data of ESO ~243-49 HLX-1. 
Previously, many properties of HLX-1 were  analyzed using $Swift$/XRT observations   (e.g., Soria et al. 2010; Farrell et al. 2013; Webb et al. 2014; Yan et al. 2015). 
In particular, Soria et al. (2010) assessed {\it Swift} (2008 -- 2009) %and $XMM-Newton$ (2004, 2008) 
observations by fitting 
%the low/hard state ($L_X\sim 2\times 10^{37}$ erg s$^{-1}$) 
their X-ray spectra. They  used  a few %four 
models, in particular, an additive model of the {blackbody/diskbb} plus power-law. % and a $diskbb$ plus $power$-$law$. 
%$zamp$ plus $power$-$law$ and $nsa$ plus $power$-$law$.   
They found that in these X-ray spectra 
%of ESO~243-49 HLX-1 
the photon index 
changes from 1.8 to 2.95 but they were unable to present any argument that  this source to be intermediate-mass BH or foreground NS. 

 \cite{{ft11}}, \cite{STSS15}, \cite{STF13}, \cite{ST12}, 
\cite{ST11}, hereafter ST11, have shown
% that one can really  distinguish  
that BH and NS sources can be distinguished 
using $\Gamma$-$\dot M$ correlation.   ST11  predicted  a BH source in 
ESO~243-49 HLX-1  and ruled out a quiescent neutron star in this source. Only in  BH sources %demonstrate that 
the photon index $\Gamma$  can increase from $\Gamma\sim 1.6$ to $\Gamma\sim 3$ with mass accretion rate, 
%so wide scatter of 
 in contrast to neutron stars,  % (NSs), 
for which  we usually find the constant
% with the constancy/stability 
 photon index around  $\Gamma\sim 2$ (see e.g. ST11). Furthermore, Wiersema et al. (2010) %, Farrell et al. (2010, ApJ, 721, L102) 
 measured a redshift of z=0.0223 for HLX-1, which clearly  indicates that HLX-1 
 cannot be a neutron star (NS) source because its luminosity is too high for a NS.   
%just as promptly concluded in ST11.

In this paper we present an analysis of  available 
{\it Swift} %and  %BEGIN NEW      $Suzaku$ and $Chandra$  
observations of ESO~243-49~HLX-1 made during  2008 -- 2015 to re-examine our previous conclusions on a BH nature of 
HLX-1 and to find further indications on intermediate-mass BH  in HLX-1.  
In Sect. 2 we present the list of observations used in our data analysis, while 
in Sect. 3 we provide  details of the X-ray spectral analysis.  We discuss the evolution of 
the X-ray spectral properties during the high-low state  transitions 
and present the results of the scaling analysis to estimate the BH mass of ESO~243-49~HLX-1 in Sect. 4. % and \S 5.  
We   conclude on  our results  in Sect. 5.

\section{Observations and data reduction \label{data}}
\begin{figure}[ptbptbptb]
\includegraphics[scale=0.95, angle=0]{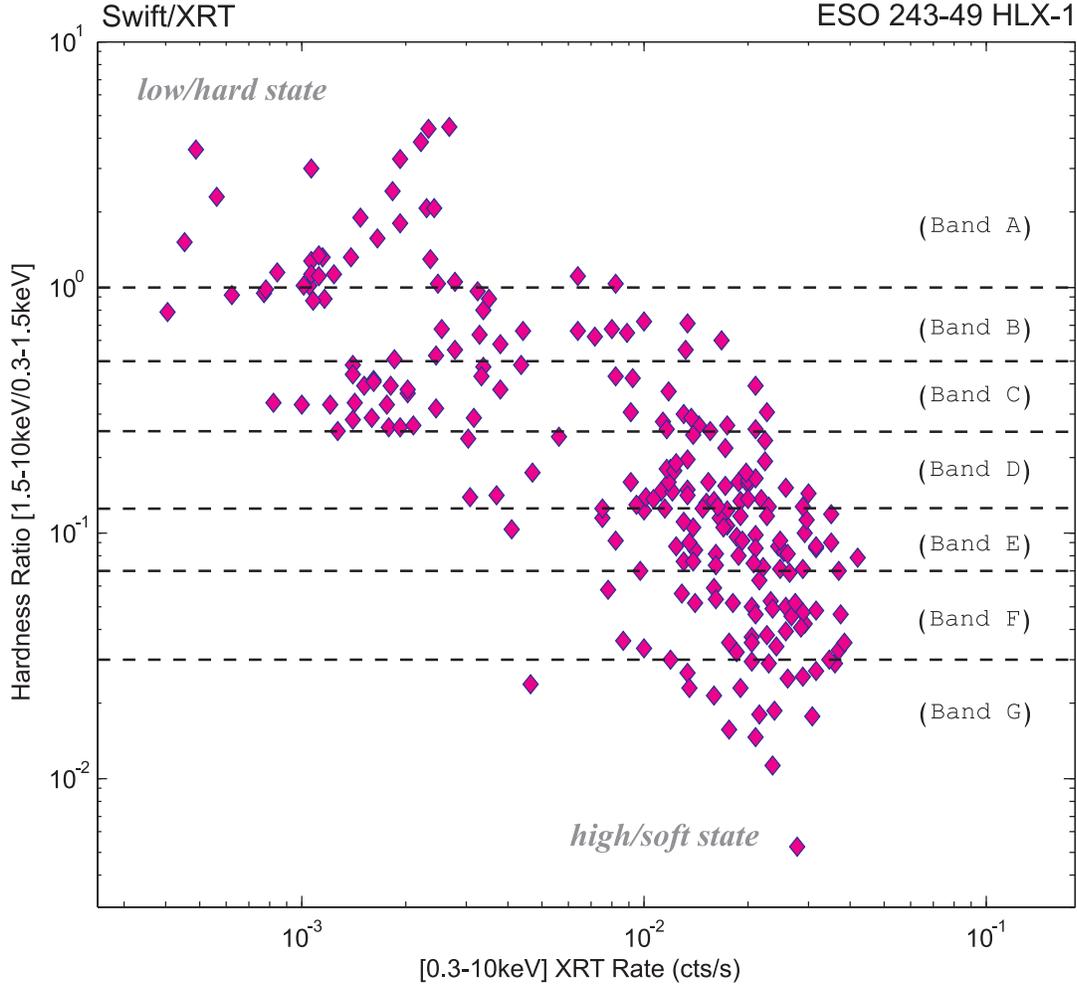}
\caption{
 Hardness-intensity %Color-intensity %evolutionary 
diagram for ESO 243-49 HLX-1 using {\it Swift} observations (2008 -- 2015) 
during spectral evolution from the low/hard state to the high/soft states. 
 In the vertical axis, the hardness ratio (HR)  is the ratio 
of the source counts in the  two bands: 
the {hard} (1.5 -- 10 keV) and  soft (0.3 -- 1.5 keV)  passbands. 
The HR  decreases with a source  brightness in the 0.3 -- 10 keV energy range (horizontal axis). For clarity, we plot only one point with error bars (shown in the bottom right corner) to demonstrate typical uncertainties  for the HR and count rate.
}
\label{HID}
\end{figure}

\begin{figure}[ptbptbptb]
\includegraphics[width=16cm]{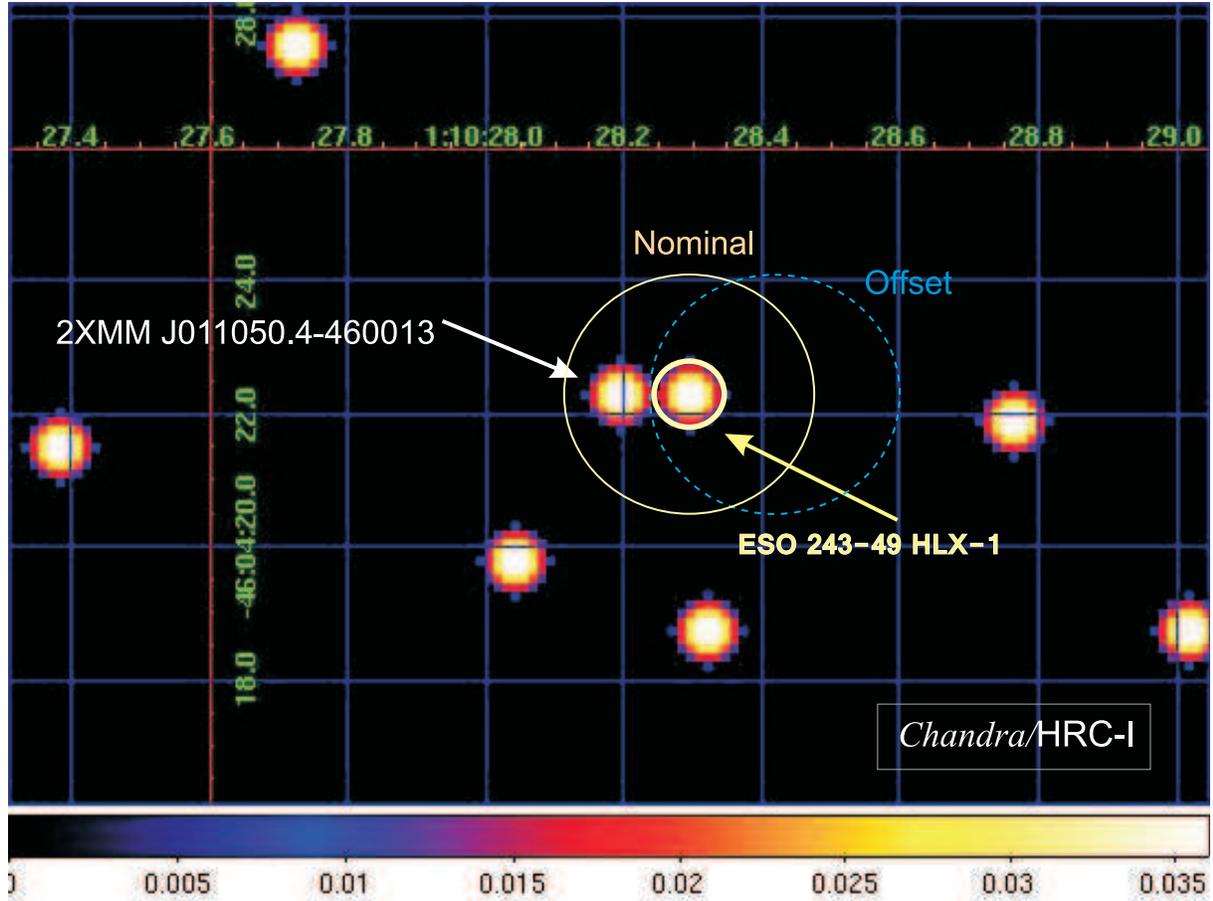}
\caption{
$Chandra$/HRC-I 
(0.1 -- 10 keV) image  of the ESO~243-49 HLX-1 field taken on UT 2009 July 4
where  yellow small circle 
corresponds to the location of ESO~243-49 HLX-1 and   white arrow points correspond 
to  2XMM J011050.4-460013 source. 
The large  circles
(labelled nominal and offset) show the two pointing positions used to extract the light curve and spectrum,   to minimize contamination of ESO~243--49 HLX--1 by a nearby source.
 }
\label{imageb}
\end{figure}

\begin{figure}[ptbptbptb]
   \includegraphics[width=17cm]{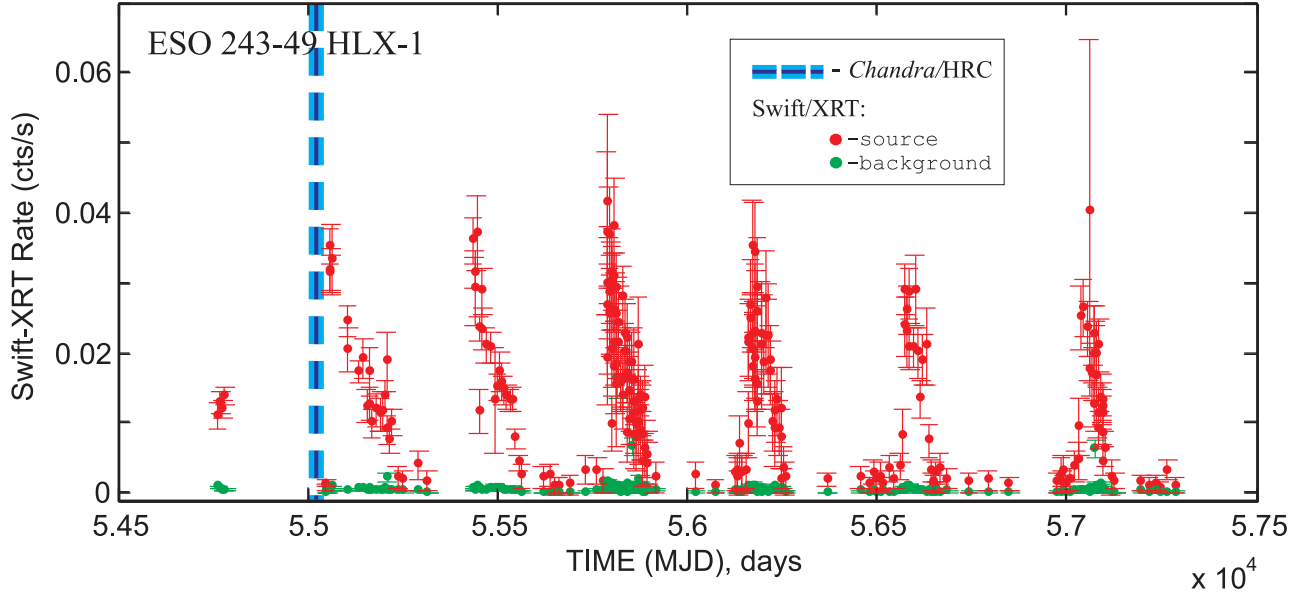}
      \caption{
$Swift$/XRT light curve of ESO 243-49 HLX-1 in the 0.3$-$10 keV energy range during 2008 -- 2015.  
Red points mark the source signal (with 2-$\sigma$ detection level)  
and green points indicate the background level.
%NEW
{
{The Blue} { dashed line}  indicates { the MJD of} the $Chandra$/HRC-I {observation} presented in Figure \ref{imageb}. 
}
% END of NEW
}
   \label{lc}
\end{figure}

We used $Swift$ data 
%{\bf
\citep{Gehrels04} %observations 
%}
of ESO~243-49 HLX-1 %M101 ULX-1 
carried out from 2008 to 2015. 
% END NEW
%}
%We performed an analysis of all $Swift$/XRT observations of M101 ULX-1 carried out from 2006 to 2013. 
In Table 1 we show the summary %log 
of the {\it Swift}/XRT 
%{\bf
\citep{Burrows05} %observations 
%}
observations. 
%{\bf
In the Swift observations,    HLX-1  has been detected,  at least, at $\sim$ 2-$\sigma$ significance 
 %, as advising in 
(see, Evans et al. 2009). 
%The {\it Swift}-XRT Photon Counting data (ObsIDs, indicated in the first column of the upper part Table 1) were processed 
%using the HEA-SOFT v6.14, the tool XRTPIPELINE v0.12.84 and the calibration files (CALDB version 4.1). 
%The {\it Swift} source count rates never exceed 0.02 count s$^{-1}$, therefore  only photon-counting mode (PC) events (selected in grades 0$-$12) 
%were considered. In this way, 
The 
%}
$Swift$-XRT/PC data (ObsIDs, indicated in the first column of %the upper part 
Table 1) 
were processed using the HEA-SOFT v6.14, the tool XRTPIPELINE v0.12.84 and the
calibration files ({latest CALDB version is 20150721\footnote{http://heasarc.gsfc.nasa.gov/docs/heasarc/caldb/swift/}}).
%We selected source events which were accumulated the grade 0$-$12 events.
The ancillary response files were created using XRTMKARF v0.6.0 and exposure maps were generated by XRTEXPOMAP v0.2.7. 
We fitted the spectrum using the response file SWXPC\-0TO12S6$\_$20010101v012.RMF.
We also applied the online XRT data product generator\footnote{http://www.swift.ac.uk/user\_objects/}  for an independent check of  
light curves and spectra (including background and ancillary response files, 
%We also obtain  images %were extracted 
%using the online XRT data processing facility\footnote{http://www.swift.ac.uk/user\_objects/} 
%(5 http://www.swift.ac.uk/user objects/) 
see Evans et al. 2007, 2009). 
%BEGIN NEW
%{\bf 
{
We also  identified  the state  using the hardness (color) ratio (HR) (see Sect.~3.2) and the Bayesian method developed  by Park et al. (2006).
Moreover,  we applied an effective area option of  the Park's code which  includes the count-rate correction factors in their calculations.  
Our results, obtained by adapting  this 
technique, indicate  
%{\bf
 a continuous distribution of the HR with source intensity from
%}
%two intensity regimes in ESO~243-49 HLX-1: i. with 
a high hardness %color 
ratio at lower count-rate  %and ii. 
%{\bf 
to
%} 
a low hardness %color 
ratio at higher count events (see Figure \ref{HID}).
Furthermore, the hardness$-$intensity diagram shows a smooth track. Therefore, 
%
%To justify a spectral state identification and data combining and we tested $Swift$ %[and $Chandra$ ?] 
%observations with a Bayesian analysis (including the background) before spectral fit procedure. Specifically, 
%in Sect.~\ref{HID_lc} we %Fig.~\ref{HID} demonstrate by way of the color-intensity diagram %, from which well seen 
%that different observations are located in different color regimes. Furthermore, the color-intensity diagram shows a smooth track. 
%Therefore we concluded that low count observations are related with low states and high count events 
%correspond to high states in M101 ULX-1.
%}
we  grouped the $Swift$ spectra into seven bands according to count rates (see Sect.~3.1) 
%: very high ("A"), high ("B"), medium ("C") and low ("D") count rates (see Fig.~2) 
and fitted the combined  spectra of each band using the 
{\tt XSPEC}{\footnote{http://heasarc.gsfc.nasa.gov/xanadu/xspec/manual/manual.html}} package 
[version 12.8.14, {see \cite{Arnaud96}}].
In addition, all  groups of the {\it Swift} spectra were binned to a minimum of
20 counts per bin in order to use $\chi^2$ statistics for our spectral fitting.
 
We also employed  {\it Chandra} data for more deep image analysis. Specifically, we investigated %a 1175 s DDT 
an observation of HLX-1 for which  the
%High Resolution
HRC-I camera was operated on the board of {\it Chandra} % on 2009 July 4 (ObsID: 10919). We extracted all the events %in the detector (i.e., the background) and plotted the number of counts per energy channel (PI) 
with  the CIAO v4.1.1 task WAVDETECT.

\section{Results \label{results}}

%\subsection{Images and light curves \label{image_lc}}
\subsection{Images \label{image_lc}}

We visually inspected %(examine) 
 the %obtained 
%{\bf 
source field-of-view (FOV)
%} 
image  to exclude  a possible contamination from nearby sources. %,  smoothed by a Gaussian with an FWHM of 3{\tt "}.
%{\bf
To do this,  we implemented   the $Chandra$ image which has a  better resolution than the  {\it Swift} image. In Figure~\ref{imageb} 
we show the adaptively smoothed $Chandra$/HRC-I %(0.06-10 keV) 
(0.1 -- 10 keV) image  of the ESO~243-49 HLX-1 field, obtained on 2009 July 4 with  an exposure of 1175 s (ObsID: 10919) 
{ when ESO~243--49 HLX--1 was in quiescence (see see also blue {dashed vertical line} in Figure~\ref{lc} which indicates  the $Chandra$/HRC-I 
observational MJD point in the lightcurve). } % Two 
Two sources were detected in the FOV near HLX--1 position: 2XMM J011050.4-460013 and 2XMM J010953.9-455538. 
 %Follow to Webb et al. (2010) and  
A source related to  the position of HLX-1 (Webb et al. 2010) is  2XMM J010953.9-455538,  identified as  ESO 243-49 HLX-1 ($\alpha=01^{h}10^{m}28^s.30$, $\delta=-46^{\circ} 04{\tt '} 22{\tt ''}.3$, J2000.0  and indicated by 
yellow circle in Figure~~\ref{imageb}). 
%Correspondingly, of which locations are indicated in Figure~\ref{imageb}  with the yellow circle (ESO~243-49 HLX-1) and the white 
%circle (2XMM J011050.4-460013).
%}
%The { Swift}/XRT (0.3 -- 10 keV) image of ESO~243-49 HLX-1 field of view 
%BEGIN NEW
%{\bf
%after smoothed by a Gaussian with an FWHM of 3"2 % (8 pixels).
%}
%END NEW
%is presented in 
%the $top$ panel of 
%Figure~\ref{imageb}. %, where  {\it green} %and $white$ circles are the locations of M101 ULX-1, NGC~5457 (M101), 
%BEGIN NEW
%NGC~5461 and M101 H~II regions. 
%END NEW
%BEGIN NEW
% Contour levels %(see the right bottom panel)  
%indicated in the $lower$ panel 
%should 
{ To separate X-ray emissions from these two sources and minimize contamination of ESO 243--49 HLX--1 by the nearby source 
we additionally used specific Swift pointing (see dashed circle in Figure~\ref{imageb}), within which only HLX--1 is detected. 
In Figure~\ref{imageb} the large circles (labeled nominal and offset) show the two pointing positions used to extract 
HLX--1 data  for the Swift observations. In this way, we compared the relative contributions of these two sources 
throughout all Swift observations. For the observations made at the 
offset pointing positions, the count rates are almost the same as  that made at the nominal pointing position. Thus,  
we conclude that during Swift observations this nearby source remains faint in comparison with the variable HLX--1.
} 

\subsection{Hardness-intensity diagrams and light curves  \label{HID_lc}}

% BEGIN NEW
%{{\bf
%Before detailed spectral fitting 
{Before we proceed with } %going to   
details of the  spectral fitting  we  study 
%investigate 
the 
%so {called} 
hardness ratio (HR). 
%in order to quantify and characterize 
%the source spectrum. 
In application to the {\it Swift} data we considered the  {HR} is a  ratio of the hard and soft counts 
in the 1.5 -- 10 keV and %$S$  and $H$ 
 0.3 -- 1.5 keV bands, respectively.
% However, at low counts, the posterior distribution of the counts ratio, $R$, tends to be skewed
%because of the Poissonian nature of data. Therefore we used the color, $C=log_{10}(S/H)$, which  a log 
%transformation of $R$, which provides the skewed distribution more symmetric (see e.g., Park et al. 2006).
%monotonic function of the counts $S$ and $H$ in 
%the soft (0.3 -- 1.5 keV) and hard (1.5 -- 10 keV) bands, respectively. Specifically,: $C=log_{10}\frac{S}{H}$. 
The HR is evaluated %modified 
by  calculating the  background counting.
% and instrumental effective areas. 
In Figure~\ref{HID}   we show  %demonstrates 
the  hardness-intensity diagram (HID)  and thus, we show  that the different count-rate observations 
 are assocated with %correspond 
 different color regimes: the higher HR  values correspond to  harder spectra.
% A hysteresis-like evolution is discernible in the color-intensity
%track in its %intermediate phase 
%middle part, which is caused by the differences between the rise and decay 
%tracks. % parameters. 
%Larger values of the {HR} indicate a , and vice versa. 
%Note, that 
% we have applied
A Bayesian approach was used  to estimate   the HR values and their errors 
%using BEHRs software %code
~\citep{Park06}\footnote{A Fortran and C-based program which calculates the 
%hardness 
ratios using the methods described by 
% the paper of 
\cite{Park06} 
%which  is available for download from 
(see http://hea-www.harvard.edu/AstroStat/BEHR/)}.
% with accounting for Poissonian nature of the observations. 
%Generally, this method is  applicable 
%when the source is faint or the background is relatively large~\citep{Evans09,Burke13,Jin06}. In our case, the most $Swift$ observations are 
%related to low count-rate regimes, which can confuse a reliable color estimates. % and % 
%However, Bayesian analysis provides a simple way to overcome  this %problem 
%problem.  
%As a  result, we demonstrate a clear low/hard to high/soft state %LS-HS 
%evolution of X-ray emission in ESO~243-49 HLX-1.
% using our data. %, including faint source signal regime 
%color-intensity evolution revealed with %show by way of 
%color-intensity diagram %so as to clear seen 
%that different count-rate observations are related to different color regimes. 
%%%
%For fainter sources we used the Bayesian method of Park
%et al. (2006), where we used the effective area option in
%their code to include the count-rate correction factors in
%the calculation. While the Bayesian method gives asymmetric errors (which are typically a few percent larger
%than the standard method returns), the standard method
%returns symmetric errors. 
%
%Furthermore, 

Figure~\ref{HID}   %clear 
indicates that the {HR} monotonically reduces  with the total count rate (0.3 -- 10 keV). %flux $S$ and achieves  
%a noticeable stability at high soft fluxes. 
%This quasi-constancy stage for the color $C$ (which is essentially proportional to 
%the spectral power-law index) a priori %additionally 
%indicates  a possible BH presence in this source (e.g., see ST09). 
%ST10, STS14).
%***
%}
%NEW
%{\bf
%Note that the color-color diagram of M101 ULX-1 clearly demonstrates two groups of datapoints,  related to the high/soft and low/hard states (see Fig. \ref{HID}). More specifically, in outbursts, M101 ULX-1 evolves from the 
%$hard$ state to the $soft$ state during the rise phase and then returned to the $hard$ state during the decay phase. 
This particular sample  is similar to those of  most of outbursts of Galactic %low mass 
X-ray binary transients (see Belloni et al. 2006; Homan et al. 2001; Shaposhnikov \& Titarchuk, 2006;  ST09; TS09; Shrader et al. 2010; Mu$\tilde n$oz-Darias et al. 2014).
%}
%END NEW

%HLX-1 has been  subjected by  
Six recurring   outbursts  occurred in HLX-1 during the {\it Swift} monitoring (from 2008 up to now). These outbursts  were  approximately separated by one year apart. % (Fig.~\ref{lc}, see also Godet et al. 2012a,b). 
%22222
 We show the {\it Swift}/XRT light curve of ESO~243-49 HLX-1 during 2008 -- 2015 for the 0.3 -- 10 keV band  in Figure~\ref{lc}. 
{Red} points indicate the source signal %(with 1-$\sigma$ detection level) 
and green points correspond to  the 
background level. %$Blue$ vertical strip marks outburst time interval (MJD 55797 -- 56057) presented 
%in the lower panel. % in more details.
We  found  six outbursts of ESO~243-49 HLX-1 peaked at MJD=55060, 55446, 55791, 56178, 56586 and 57060 
with a FRED 
%(``fast rise exponential decay'') 
profile and rough duration from 70 to 200 days. For 
the remaining  {\it Swift} observations the source were in the {low state}. 
Unfortunately, individual {\it Swift}/XRT 
observations of ESO~243-49 HLX-1 in  photon counting (PC) 
mode do not have enough counts in order to make  statistically significant spectral fits. 

%To solve %get around 
%this problem, 
We studied   the {\it Swift}/XRT HID (see Figure~\ref{HID}) and grouped the 
observations into seven bands: very low (A, HR$>$1; B, 0.5$<$HR<1), low (C, 0.25$<HR<$0.6), 
intermediate (D, 0.13$<HR<$0.25), high (E, 0.07$<HR<$0.13; F, 0.03$<HR<$0.07) and 
very high (G, HR$<$0.03) count rates to resolve this difficult problem. 
%("A" [$HR>1$], "B" [$0.5<HR<1$]), low ("C" [$0.25<HR<0.6$]), intermediate ("D" [$0.13<HR<0.25$]), 
%high ("E" [$0.07<HR<0.13$], "F" [$0.03<HR<0.07$]) and very high ("G" [$HR<0.03$]) count rates (see Fig.~\ref{HID}). 
%very high ("A"), high ("B"), medium ("C") and low ("D") count rates (see Fig.~\ref{lc}). 
%NEW
%{\bf
%We have also  split Band C into two subbands. Blue points shown in Figure~\ref{HID}  are associated with 
%softer/higher track (see also related points in  
%and also seen
% an averaged 
%the  lightcurve, Fig.~\ref{lc}). %4). 
%In fact, this softer track ($blue$ points of Figure~\ref{HID}) corresponds to the outburst 
%decay part (%marked by {\it blue} color 
%see Fig.~\ref{lc}). 
%This  fact allows 
%to suggest 
%It is possible that these  two tracks can be caused by differences between the rise and decay  conditions. This allow us to 
%We have divided Band-C into two subbands: Band-C$_s$ formed by $blue$ points of the softer/upper track in  the color-intensity diagram and 
%While Band-C$_h$ (red points) are 
%formed by other points of Band-C 
%related to  the lower track of  the color-intensity diagram.
%}
%END of NEW
To do this, we  have combined %combined 
the spectra  in each  related band, regrouping  them with the task grppha and then
fitted %analysed 
them using the 0.3 -- 10 keV  range. %with the XSPEC task, 
%using %both 
%the Cash %and $\chi^2$ statistics.
% and fitted them for all these observations
%{\bf 
%using $\chi^2$ statistics.
%In addition, some of the brightest source spectra of A- and B-sets %groups 
%were regrouped with the task grppha and then
%analysed in the 0.3 -- 7 keV  range %with the XSPEC task, 
%using %both 
%the Cash %and $\chi^2$ 
%statistics.
%}

% such produced spectra.
% of each band.

%Consequently, we %grouped the $Swift$ spectra into four bands according to count rates (Sect.~\ref{results})  and 
%fitted the combined spectra of each band with {\tt XSPEC} package (version 12.8.14).

\subsection{Spectral Analysis \label{spectral analysis}}

Different spectral models were used  in order to test them  for all available data  for ESO~243-49 HLX-1.
%with the aim 
We wish to establish  the low/hard and high/soft state evolution using  these spectral models.
% in frame of the same model. 
We investigated  the combined $Swift$ %, $Suzaku$ and $Chandra$ 
spectra 
%of M101 ULX-1 
%related to  different spectral states (the HR bands, see Fig.~\ref{HID}) 
to test  the following spectral models: 
powerlaw, blackbody, BMC and their possible combinations modified by an absorption  model. 
%NEW
{
%We have also analysed %investigated 
%a possible absorption effect in X-ray spectra of M101 ULX-1 
%(from neutral matter) 
%taking into account 
%considering two possible cases for the absorbing column: 
%($i$) the constant  $N_H$ %along state transition 
%and 
%($ii$) the variable $N_H$ %ranging from 0.1 to $5\times 10^{21}$ cm$^{-2}$ 
%during the state transition. 
}
 To  fit all of these spectra, we used a  neutral column, which was obtained by  the best-fit  column $N_H$ 
%{\bf 
of $5\times 10^{20}$ cm$^{-2}$ 
%}
(see also Yan et al. 2015; Farrell et al. 2009; Webb et al. 2010, 2012). 
%So we fixed the column density at the line-of-sight value $N_H=5\times 10^{20} cm$^{-2}$ for all the three spectra.
%}

%In particular, we  try to find the absorption effect %contribution of 
%  based on the $Swift$
%  and $Chandra$ spectrum samples 
% NEW
%{\bf 
%for two cases: ($i$) the constancy of $N_H$ along the state transitions 
%and ($ii$) the variability of $N_H$ along the state transitions. 
%}
%(Sect.~\ref)
\subsubsection{Choice of the Spectral Model\label{model choice}}

%As a first step, we proceed with a model of an absorbed power-law.  
The phabs*power-law model  fits   the low state data well [e.g., for band-A spectrum, $\chi^2_{red}$=1.15 (138 d.o.f.), see 
the  top of Table~2]. As we established   this  power-law model indicates to  very large photon 
indices (greater than 3, particularly for band-G spectrum, see Figure \ref{HID}) 
%in each case and 
and moreover, this model  has unacceptable fit quality, $\chi^2$  
%for the high state data 
 for all D, E, F and G-spectra of $Swift$ data. 
%on 2004 July, December 30 and January 1 of $Chandra$ data). 
For the high-state data, the thermal model (blackbody) provides better fits 
than the power-law model. However, the intermediate state spectra  (D-spectra) % for $Swift$ data)
% and $Suzaku$ data) spectra 
cannot be fitted by any single-component model. A simple power-law model produces a soft excess.  These significant positive residuals at low energies, lower than 1 keV, suggest the presence of additional emission  components in the spectrum. 
As a result  we also tested a sum of blackbody and power-law component model. The model parameters  are  $N_H=5\times 10^{20}$ cm$^{-2}$; 
$kT_{bb}=90-300$ eV and $\Gamma=1.3 - 2.9$ (see more 
%NEW
%{ the left column  of }
in Table~\ref{tab:par_swift}). % 2).
% the power-law component contributes roughly 60\%  and 25\%  of 
%the absorbed  and unabsorbed 0.3 -- 7 keV flux respectively. 
The best fits of the {\it Swift} spectra has been found  using of  the  
{bulk motion Comptonization model} [{BMC XSPEC} model, \cite{tl97}],   
% the {\it Bulk Motion Comptonization} 
%(BMC) model 
for which $\Gamma$ ranges from 1.6  to  3.0  for all observations (see Table~\ref{tab:par_swift} % 2 
and Figures~\ref{6_swift_sp_compar}-\ref{two_state_spectra}). %two_state_spectra}). Furthermore, we achieve the 
We emphasize that all these best-fit results are found using the same model for 
%all spectral 
the high and low states. 

\subsubsection{Spectral modelling for ESO~243-49 HLX-1\label{bmc-results}}

%{\bf
Now, we  briefly recall  the physical picture described  by the $BMC$ model, its key assumptions and  parameters. 
The BMC   Comptonization spectrum is  a sum of a part of the blackbody (BB) directly visible by the Earth observer [a fraction of $1/(1+A)$] and a fraction of  the BB, $f=A/{1+A}$, convolved with the Comptonization Green function which is, in  the BMC approximation,  a broken power-law. It is worthwhile to emphasize that this  Comptonization Green function is  characterized by  only one parameter, the spectral index $\alpha=\Gamma-1$.  
Thus, as one can see that the BMC model has the main parameters, $\alpha$, $A$, the seed blackbody temperature $T_s$ and the BB normalization which is proportional to the seed blackbody luminosity and inverse proportional to $d^2$ where d is a distance to the source   (see Figure~\ref{geometry}).

 The spectral evolution  for the low/hard state-high/soft state 
%we collect in the summary %general 
%picture of 
(LHS$-$HSS)  transition is evident  in Figure~\ref{6_swift_sp_compar}.  Seven %six representative 
$EF_E$ spectral diagrams related to different spectral states in HLX$-$1 are presented there
(see  also Sect.\ref{HID_lc} and  Figure \ref{HID}).

Thus, the BMC model  successfully fits  the ESO~243-49 HLX-1 spectra
for all spectral states. In particular, %Therefore, ULXs are considered to commonly have TCS coro-
the $Swift$/XRT spectra for band A (red) and band G (blue) fitted using the BMC %$bmc$ 
model are shown in Figure~\ref{two_state_spectra}. % ($left$ panel). 
%Note,  this strong  spectral variability between the sets of the observations is also seen in the hardeness intensity diagram (HID), see Figure~\ref{HID}. 
%for our definition of $Swift$/XRT hardness ratio bands, and Table~\ref{tab:par_swift} % 3 
%for the best-fit parameters). 
In Table~\ref{tab:par_swift} %3 
(at the bottom), we present the results of spectral fitting  $Swift$/XRT data of ESO~243-49 HLX-1 using our main spectral model. 
phabs*bmc. % for a particular case of $N_H=5\times 10^{21}$ cm$^{-2}$ (see the $left$ hand side  of 
%Table~\ref{tab:par_swift}). % 3). 
In particular, the LHS$-$HS %$low-high$ state 
transition  
%M101 ULX-1 
%based on $Swift$/XRT data 
is related to 
%the spectral parameter evolution, %:  
 the photon index, $\Gamma$ changes from 1.6 %$1.4<\Gamma <2.5$, 
to 3.0 when the relatively low seed photon temperature $kT_s$ changes from 50 eV to 140 eV. 
The BMC normalization, $N_{bmc}$ varies by  a factor of fifteen,  in the range of 
$0.3<N_{BMC}<5.2\times L_{33}/d^2_{10}$ erg s$^{-1}$ kpc$^{-2}$,  %)
%{for a constant $N_H$ case and by factor twenty ($0.6<N_{BMC}<12.6\times L_{35}/d^2_{10}$ erg s$^{-1}$ kpc$^{-2}$) 
%for a variable $N_H$ case. 
while
%and 
the Comptonized (illumination) fraction varies in a wide range %about an order of magnitude 
($-1.2<\log{A}<1.1$ or $f\sim 0.1-1$). % quite low ($\log{A}<-4$ or $f\sim 10^{-4}$) {for all cases.}
%{\bf
%We also fitted some individual Swift spectra of A- and B-groups using Cash-statistics, which allows better investigate soft 
%state evolution of M101 ULX-1 at its bright events. In particular, these spectra better provide plateau part 
%(well seen in Fig.~8) of the $\Gamma$ -- $\dot M$ %index-mass accretion rate 
%track, where index saturation effect is expected.
%}

%As we wrote above examples of the best-fit spectra of ESO 243-49 HLX-1 are presented in 
In Figure~\ref{two_state_spectra}  we  also plot the spectral evolution of ESO 243-49 HLX-1
in E*F(E) units (top) along with $\Delta\chi$ (bottom). Data are taken from $Swift$/XRT observations with  
HR=$1.1\pm 0.1$ [red, LHS; $\Gamma=1.6\pm 0.2$, $T_s=54\pm 9$ eV, $\chi^2_{red}=0.86$ (136 d.o.f.)] %, $HR=0.15\pm 0.03$ ($central$, IS),
and HR=$0.02\pm 0.01$ [blue, HSS; $\Gamma=3.0\pm 0.1$, $T_s=130\pm 10$ eV, $\chi^2_{red}=1.04$ (217 d.o.f.), 
see also Table~\ref{tab:par_swift} %2 
for details]. %Here data are denoted by $black$ points; the spectral model 

{ 
In the fit of the HSS spectrum  shown in the third panel of Fig.~\ref{two_state_spectra}, the BMC model appears to be systematically  over-predicting the strength of the thermal emission below 1 keV.  This energy range is related to the oxygen and 
iron line region where  absorption features occur at 0.6 keV and 0.9 keV,  most likely associated with O~VIII Ly$\alpha$ and Fe~XVIII -- Fe~XIX. 
%Note, that these features were established by Godet et al. (2012a) using {\it Swift+XMM-Newton} spectra of HLX--1 in the framework of  another continuum model as {an absorbed diskbb+powerlaw}. 
However, this possible complexity of the model is not well constrained by our data.
}

As we have already discussed  above, the spectral evolution of ESO~243-49 HLX-1 was previously investigated  using the $Swift$ data by many  authors. In particular, Soria et al. (2010) and Yan et al. (2015) studied the 2008 -- 2009 and 2009 -- 2015 $Swift$-data sets (see also Table 1) using an additive diskbb plus power$-$law model and  a simple power$-$law model. %%$wabs*(diskbb+powerlaw)$ model and $wabs*power$-$law$ model 
%(for 2008 -- 2009 and 2009 -- 2015 data sets), 
These  qualitative models   describe an evolution of these spectral model parameters throughout % and confirmed the presence of spectral 
state transitions during the outbursts. 

We have also found a similar 
spectral behavior using our model and the full set of the $Swift$ observations.  
In particular, as Soria's and Yan's et al., we also found that % ESO~243-49 
%HLX-1 demonstrates monotonic growth of
 the photon index $\Gamma$ of HLX-1 grows monotonically
during the LHS -- HSS transition from $\sim$1.6 to 3. In addition, we found that  $\Gamma$ tends to  saturate  at  3 
%the %an upper level of $\Gamma=3$ 
at high values of $N_{bmc}$. In other words  we  found  that  $\Gamma$  saturates at high values of  mass accretion rate. 

   Seed photons  with the lower $kT_{s}$, %presumably 
related to a lower mass accretion rate,  
are Comptonized more efficiently in the LHS because, we revealed that the illumination fraction $f$ [or  $\log(A)$] is quite high in this state. 
% seed photons can be Comptonized more efficiently up to higher energies,
%which ends up with broad continua. 
In contrast, %In contrast, %On the other hand 
in the HSS, these parameters, $kT_{s}$ and $\log(A)$  show an opposite behavior, namely 
 $\log(A)$ is lower for higher $kT_s$.  This means that 
%In this state 
a relatively small fraction of the seed photons, whose  temperature is higher because of the higher mass accretion rate in the HSS than that in the LHS, is  Comptonized.
%Therefore, seed photons can not be Comptonized effciently and form the
%round spectral shape.
%pectra are described by a $phabs*bmc$ model  with $kT_{bb}=70$ eV ($red$), for  the high state and  with 
%kT_{bb}=45$ eV ($blue$), for the  low state. See also the best-fit parameters %correspond to the fits 
%isted in Tables~\ref{tab:par_swift} and \ref{tab:fit_table_Chandra} for $Swift$ and $Chandra$ data, respectively.
%Similarly to $Swift$ data analysis, we have also analyzed  a case of variable $N_H$ during the LS-HS transitions and found that
%the $BMC$ normalization varies by  a factor of seventeen ($2<N_{BMC}<35\times L_{35}/d^2_{10}$ erg s$^{-1}$ kpc$^{-2}$) 
%for a constant $N_H$ case and by a factor of sixty ($0.5<N_{BMC}<38\times L_{35}/d^2_{10}$ erg s$^{-1}$ kpc$^{-2}$) 
%for a variable $N_H$ case (see $right$ colunm part of Table 4). While the Comptonized (illumination) fraction is again quite low ($\log{A}<-4$) for all cases.

Our spectral model performs  very well  throughout
all data sets. %(187 observations) 
{In Table~\ref{tab:par_swift} % 3 
we list  a good performance of the BMC model when it is applied to the $Swift$ data ($0.79<\chi^2_{red}<1.14$).}
The reduced  $\chi^2_{red}=\chi^2/N_{dof}$ %(see  examples of the spectral fits in Table~\ref{tab:par_swift}) % 2)
%used in our analysis. Namely, a value of reduced
(where $N_{dof}$ is the number of degree of freedom) is  lower than  or around  1 %1.0 
for  all %the most of  the 
observations. 
%For a small fraction (less than 3\%) of the spectra with high counting statistics
%$\chi^2_{red}$ reaches 1.4. However, it never exceeds a rejection limit of 1.5. 

We  also estimate the radius of the blackbody emission region.
% as a suitable equivalent to the Bbody normalization, defined above. 
We  found   the blackbody radius $R_{BB}$ derived using a relation 
$L_{BB} = 4\pi R^2_{BB}\sigma T^4_{BB}$, where $L_{BB}$ is the seed blackbody  luminosity  and $\sigma$ is  the Stefan-Boltzmann constant. With  a distance D to the source  of 95 Mpc, we obtain that the region associated with the blackbody has  a radius of 
$R_{BB}\sim 5\times 10^6$ km.  
Such a large BB region is only  around   the IMBH  which means that ESO~243-49 HLX-1 probably   is the IMBH source.  
%{\bf 
Taking into account that the BMC normalization varies by a factor of 15 and  the
blackbody (seed) temperature $T_{s}$ changes by a factor of 2 (see Table 2) we find that the blackbody radius is almost constant. 
%}
%in frame of our approach.
%the bbody radius changes by a factor of four ($1.3<R_{BB}< 5\times 10^6$ km).  
%In comparison, 
%}
We note that
 $R_{BB}$ is of order of $10-30$ km for a Galactic BH with a mass of around  10 solar masses (see STS14).   

%\subsubsection{Evolution of X-ray spectral properties during spectral state transitions}

%13131
 We have also found that  the emergent spectra of ESO~243-49 HLX-1  undergo an evolution from   %characteristics of 
the low/hard state  to the  high/soft 
%(rise-decay) 
  state (see Figs.~\ref{HID}  and \ref{6_swift_sp_compar}). %transitions 
%of ESO~243-49 HLX-1 
%(as they are seen in Figs.~\ref{HID} and \ref{lc}).  
%based on their
%spectral parameter evolution of X-ray emission in the energy range from 0.3 to 10 keV using $Swift$/XRT. %, $Suzaku$ and $Chandra$/ACIS data. 
%In Figure~\ref{two_state_spectra} %$-$\ref{two_sp_Chandra} 
%we present typical examples of the 
%$Swift$ %, $Suzaku$ 
%and $Chandra$ spectra for LS$-$HS states of M101 ULX-1, while 
%In Figure~\ref{lc} 
%and \ref{chandra_lc} 
%we show the light curves outlining %demonstrating 
%the X-ray variability of the source. 
In Table~\ref{tab:par_swift} %Figure~\ref{chandra_lc}, {\it bmc model case} % from top to bottom} 
we present  the change in  %{\it RXTE}/ASM count rate, 
%model luminosity $L_x$ in 0.3$-$7 keV %10 keV % and 10-60 keV 
%energy range,  %s ({\it blue and crimson} points respectively), electron 
the seed photon temperature $kT_s$, % (in eV),  
the BMC %and $blackbody$ 
normalization %s  ({\it crimson, blue} respectively)  
and  the photon index $\Gamma$ during 2008$-$2015 outburst transitions observed with $Swift$/XRT. %set ({\it R4}). 
%The outburst %rising 
%phases  of the LS$-$HS %$mild$ 
%transitions are marked by blue vertical strips. 

In particular,    we detected   a high value of the seed photon 
temperature $kT_{s}=130$ eV  in the high/soft state (see Table~2, band-G). %in e.g. MJD=53194 point) 
%along with the maximum 
%and efficiency of the plazma Comptonization 
%increases along with 
%of the normalization $N_{bmc}$. %drops to its maximal %minimal value. 
On the other hand, the  {low/hard state}  of ESO~243-49 HLX-1 is associated with the low  $kT_{s}\sim 54$ eV (see Table~\ref{tab:par_swift}, band-A).  %(see e.g. MJD=53000 -- 53150 interval in $T_s$-panel of Fig.~\ref{chandra_lc}) 
%and the low %higher 
%Comptonized fraction $f$ [$f = A/(1 + A)$] % [which is related with $\log(A)$, $f=A/(1+A)$] %$f$ 
%of the BMC component 
%(see Table~\ref{tab:par_swift}). %-$\ref{tab:fit_table_Chandra}).
% when 
%the photons are effectively downscattered to the lower spectrum end, as expected according to our model.
%While, during the rise of the outburst, the spectra were supersoft with the temperatures 
%of seed photons $kT_s$ of 40 -- 70 eV, the photon index $\Gamma$.

%From this Table clearly seen that all spectral parameters correlate with each other during  the LHS$-$HSS transitions. 
We also establish that 
%the correlations of 
the photon index $\Gamma$ correlates with the  BMC normalization,  $N_{BMC}$ %in units of $L_{39}/D^2_{10}$. 
(proportional to mass accretion rate $\dot M$) and finaly saturates at high values of $\dot M$ (see Figure~\ref{saturation}). %, where 
%In Figure \ref{saturation} 
%where one  can see that  
%The best-fit parameters are presented as diagrams (Figure 3), where clear seen that the photon index 
The  index $\Gamma$
 monotonically grows from 1.3 to 2.8 with $\dot M$
%$N_{BMC}$ (proportional 
%to the  disk mass accretion rate 
%to $\dot M$) 
and then finally saturates  at $\Gamma_{sat}=3.0\pm 0.1$ for  high values of $\dot M$.
% $N_{BMC}$ (or mass accretion rate).
%One can see the strong saturation effect of the index $\Gamma$ versus  $N_{BMC}$. %, particularly for 

%POTOM  !!!!!!!!!!!!!!!!!!! low state
%{\bf 
%Assuming %Fixing 
%the $N_H$ at the $5\times 10^{21}$ cm$^{-2}$ and obtained the 0.3 -- 7 keV luminosity is about $(1.5 - 4.6)\times 10^{37}$ 
%ergs s$^{-1}$ (assuming 6.7 Mpc as a distance to ULX-1). Fixing the $N_H$ at a lower value ($10^{21}$ cm$^{-2}$; see below), 
%the luminosity is about a factor of 2 lower.
%}
%END NEW

}
%\section{Results and Discussion \label{disc}}
\section{Discussion \label{disc}}

%Before  proceeding with the interpretation of the observations,
%let us briefly summarize them as follows. 
%(1) The spectral data of ESO~243-49 HLX-1 are well fitted by the BMC model for all 
%analyzed LHS and HSS spectra (see Figures~\ref{6_swift_sp_compar}-\ref{two_state_spectra} 
%and  Table~\ref{tab:par_swift}). %-$\ref{tab:fit_table_Chandra}]. 
%(2) The presence of a neutral absorber with $N_H\ge 5\times 10^{21}$ cm$^{-2}$ is 
%required to explain the observed curvature of the spectrum below $\sim$0.5 keV. 
%[see, for example Figure~\ref{suzaku_spectra_N_H}]. 
%(3) 
%(2) The Green's function index of the BMC component $\alpha$ (or the photon index $\Gamma=\alpha+1$) 
%rises and saturates with an increase of the BMC normalization (proportional to $\dot M$). The photon 
%index saturation level of the BMC component is about 3.0 (see Figure~\ref{saturation}). 
%}
%}
\subsection{Saturation of the  index as a  signature of a BH  \label{constancy}}

After applying %Using 
our analysis of the evolution of the photon index $\Gamma$  in ESO~243-49 HLX-1 we  probably find 
%we have firmly using  {\it Swift} observations 
%we have firmly established 
 the photon index, $\Gamma$ saturation  with  mass accretion rate, $\dot M$.
%(in fact, the BMC-normalization $N_{BMC}$,  which is proportional
% to the (disk) mass accretion rate, 
%to $\dot M$}. Thus  ST09 provided %provide 
 ST09 have reported  that this index saturation is a first indication of the converging flow into a BH. 

\cite{tlm98} have reported  using the equation of motion that 
%the transition layer (TL), 
the innermost part  of the accretion flow (TL)  shrinks  
%(getting more compact) 
when 
%mass accretion rate, 
$\dot M$ grows.  
 It is worthwhile to emphasize that  for a BH the photon index $\Gamma$  grows and  saturates 
%emergent spectrum saturates, 
for high 
%mass accretion rates 
$\dot M$.  \cite{tz98}, hereafter TZ98,  semi-analytically discovered  the saturation effect  and later  \cite{LT99}, (2011), hereafter LT99 and LT11, 
% \cite{LT11} 
confirmed this effect   making Monte Carlo simulations. 
% $\dot M$ increases. 
%Analyzing a number of   ST09, \cite{tsei09}, \cite{ST10} and \cite{STS14} (STS14) confirm the  LT99-11 prediction   
%analyzing quite a few BH candidate source  that 
%the photon  

Observations of many  Galactic BHs (GBHs) and their X-ray spectral analysis 
[see  ST09, \cite{tsei09}, \cite{ST10} and STS14]
have  confirmed    this   prediction of TZ98.
%In  Figure~\ref{saturation}  
For our particular source HLX-1,  we also found  that   $\Gamma$ 
monotonically increased  from 1.6  and then they  finally saturated at a value of 3.0  (see  Figure~\ref{saturation}).  
%(see Fig. \ref{saturation}). 

Using the  index-$\dot M$ correlation found in ESO~243-49 HLX-1   we  estimate a BH mass  in this source by scaling this correlation with those   detected  
in a  number of GBHs and M101 ULX-1 (see details below, in Sect. 4.3).

\subsection{X-ray spectra of ULXs}

{ As we have   pointed out  
%in the Introduction, 
above,  there are different scenarios for a ULX central 
source: stellar mass black hole, intermediate-mass black hole and neutron star.
% (e.g., M82 X--4). 
The ULX population may not not be homogeneous, and therefore  different ULXs may be related to different origins.  In particular, Soria \& Kong (2016) developed arguments to introduce  two sub-classes of ULXs: 
%which are still hard to explain: 
(i) hyperluminous X-ray sources (e.g., ESO 243--49
HLX--1) 
and supersoft ULXs (e.g., M101 ULX--1), but discussion of this classification  is beyond the scope  of  this  paper.
% see more detailes in Soria \& Kong (2016) and Di Stefano \& Kong (2003)).
However, in spite of these differences and their classifications as ULXs, 
%in terms of the index-mass accretion rate correlations 
we can suggest a possible similarity between 
ESO 243--49 HLX--1 source and  M101 ULX--1 and a number of GBHs in terms of their index-mass accretion rate ($\Gamma$-$N_{bmc}$) correlations. 
%based on our analysis of X-ray spectral data.
}   

%Our analysis of X-ray spectral data   from ESO 243-49 HLX-1 reported in this Paper indicates  to similarity of this source with  M101 ULX-1 and a number of GBHs in terms of the index-mass accretion rate correlations.
% revealed for these sources. 
%The X-ray observations  indicate that HLX-1 more resembles GBHBs scaled to
%higher luminosities %(three orders of magnitude) 
%than any other category of ULXs, for example, . 

{We note that   Yan et al. (2015) did not} find any similarity  between 
HLX-1 and GBHBs in terms of  the relations of the  total radiated energy versus peak luminosity, as well as
the total radiated energy vs. e-folding rise/decay timescales (see Figures 8 and 10, respectively  there). 
%Thus, it is important for us  to compare 
%the ULXs in terms of the $\Gamma$-$\dot M$ correlation  and to check the self-consistency of our approach in application to ULXs. 
On the other hand,  we compared the $\Gamma$-$N_{bmc}$ 
%(mass accretion rate) 
correlations of M101 ULX-1,  ESO~243-49 HLX-1 and those found in GBHs, { and found they are self-similar.  We therefore  applied them}   to estimate   BH mass in ESO~243-49 HLX-1 (see Figure~\ref{three_scal}).

\subsection{Estimate of BH mass in ESO~243-49 HLX-1 and  comparison with the estimate found in the literature \label{bh_mass}}

%% 111
To estimate BH mass, $M_{BH}$ of   ESO~243--49 HLX--1, we chose three galactic sources [XTE~J1550--564, H~1742--322 (see ST09) and 4U~1630-47 (see STS14)] and the extragalactic source M101 ULX-1 (see TS16),   whose BH masses and  distances  have been  estimated previously  % established now 
(see Table~\ref{tab:par_scal}), as the reference sources.  
{In particular, the BH mass  for  XTE~J1550--564 was estimated by dynamical methods.}  
For the BH mass estimate { of   ESO~243--49 HLX--1} we  also used  the BMC normalizations, $N_{BMC}$ of these reference sources.  %and  index-QPO frequency  
%patterns previously  found  in ) 
Thus, we scaled  the index vs  $N_{BMC}$  correlations for these reference sources  with that of 
the target source   ESO~243-49 HLX-1 (see Figure~\ref{three_scal}). 
The value of the  index saturation  is   almost the same, $\Gamma\sim3$  for all these target and reference sources.   We applied the correlations found in  these four reference sources to   
 comprehensively cross-check of  a BH mass estimate for ESO~243-49 HLX-1.
% We  have used %at least two 
%{a number of} 
%these four reference sources  for the comprehensive cross-check of a BH mass evaluation of ESO~243-49 HLX-1.

Figure \ref{three_scal} shows the correlations of the target source (ESO 243-49 HLX-1) and the reference sources   have  similar shapes and index saturation levels.
{This}  allows us to make 
% showing the same index saturation level, which allows us to carry out %perform 
a reliable scaling of these correlations with that of  ESO 243-49 HLX-1.
% and 
%the similar shapes for both source tracks. 
%Thus   these two values of the scaling coefficients can provide us a BH mass estimate for 4U~1630--47.
%We can proceed with this 
%scaling if these two correlations  are self-similar. Then the value of the scaling coefficient provides us 
%BH mass estimate. Note that the index-normalization correlation curve  for  1998 rise data 
%of XTE~J1550-564 (taken from ST09) is self-similar with that we find  for 4U~1630--47.
To implement the   scaling technique, we introduce an analytical %{\bf fitting} 
approximation  %generally based on the parameterization 
of the $\Gamma-N_{bmc}$ correlation, %{\bf approximated} 
fitted 
by a function (see also ST09)
 %according to ST09 

%ffffff
%\end{document}

\begin{equation}
F(x)= {\cal A} - ({\cal D}\cdot {\cal B})\ln\{\exp[(1.0 - (x/x_{tr})^{\beta})/{\cal D}] + 1\}.
\label{scaling function}
\end{equation}
with $x=N_{bmc}$.

%ffffff
%\end{document}

%As a result of 
As a result of fitting   the  observed correlation  by  this function $F(x)$
%expression %form 
%to the correlation pattern, we determine %find 
we obtained a set of the best-fit parameters $\cal A$, $\cal B$, $\cal D$, $N_{tr}$, and $\beta$.  
%that represent a best-fit form of the function $F(x)$ for a particular correlation curve. 
%The meaning of the parameters is the following: 
The meaning of these parameters is  described in details in our previous paper [Titarchuk \& Seifina (2016), hereafter TS16].
%For $x\gg x_{tr}$, the function $F(x)$ goes to a 
% value $\cal A$ which 
%is the value of the index saturation level, $\beta$ is the power-law index 
%of the curve $F(x)$ and $x_{tr}$ is a value of $N_{bmc}$ at which the index $\Gamma$ starts 
%to grow, while $\beta$ provides  the slope of the correlation. 
%The parameter $\cal D$ determines  a smoothness of  the  function when   $F(x)$ saturates to 
%$\cal A$. 
This function $F(x)$ is widely used for a description 
of the correlation of $\Gamma$ versus $N_{bmc}$ %or $\Gamma$ vs $x$ 
[\cite{sp09}, ST09, \cite{ST10}, \cite{STS10}, STS14, \cite{ggt14} and TS16]. 

%We scale the data to a template by applying a transform $N\to s_N\times N$ until the best fit is found. 
%(see Fig.~{three_scal}). 

To implement this BH mass estimate for the target source, we  relied on 
the same shape of the $\Gamma-N_{bmc}$ correlations for the target source and those for the reference sources.  
%A key %crucial 
%assumption for this technique to be applied is that different reference sources show  and 
The only difference in values of $N_{bmc}$  for these four sources is   in 
   a ratio of BH mass to the squared distance, 
%namely in the coefficient 
$M_{BH}/d^2$. 
 Figure ~\ref{three_scal}  shows    
  the index  saturation value, $\cal A$  is approximately the same for the target and reference sources (see also  the second column in Table 3). %(see {\it bright blue} vertical arrow) 
%is 
%the value of the index saturation level, 
%almost the same for all scaling sources. In  other words, the best-fit parameter $A$
%(within the limits of error bars)  is almost the same for all these sources. 
%Thus, to justify the fact that all scaling sources 
%M101~ULX-1, XTE ~J1550-564, 4U~1630-472 and H~1743-322 
%have the same index saturation level, it should simply compare their 
%``A''-parameters . 
%In particular, $\cal A_{HLX}=3.0\pm 0.1$, $\cal A_{ULX}=2.8\pm 0.1$, $\cal A_{1550}=2.84\pm 0.08$ 
%and $\cal A_{1743}=2.97\pm 0.07$ for ESO~243-49 HLX-1, M101 ULX-1, XTE J1550-564 and H 1743-322, respectively.
% As is evident from the foregoing these ``A''-parameters (per se, saturation 
%levels) are the same within the limits of error. 
For example,  %the black  horizontal arrow stresses that 
 the shapes of the correlations for  ESO~243-49 HLX-1 (black line) and H~1734-322 % XTE~J1550-564 
(green line)]  
 are similar and the only difference of these correlations  
%  to be have the same way, the
   is in the BMC normalization values (proportional to 
% because of   
%being the horizontal interval here in 
  $M_{BH}/d^2$ ratio).
  
%Thus, in order to derive % obtain
To estimate BH mass,  $M_t$  of ESO~243-49 HLX-1 (target source) we shifted the reference source correlation along $N_{bmc}-$axis  to that of the target source (see Fig. \ref{three_scal}).

\begin{equation}
M_t=M_r \frac{N_t}{N_r}
\left(\frac{d_t}{d_r}
\right)^2 f_G,
\label{scaling coefficient}
\end{equation}

\noindent where t and r correspond to the target reference sources, repectively
% r stands for the reference  sources  
and a  geometric factor,  
$f_G=(\cos\theta)_r/(\cos\theta)_t$, the inclination angles $\theta_r$,  
$\theta_t$ and $d_r$, $d_t$ are distances to the reference and target sources respectively (see ST09). %Here subscripts $r$ and 
%The geometrical factor $f_G$ takes into account the accretion geometry, particularly for the disk geometry,  while $f_G\sim1$ in the case of  spherical accretion 
The values of $\theta$ are listed in  
Table \ref{tab:par_scal} and when some of these $\theta$-values  were unavailable then we assumed that %$\theta_t\sim \theta _r$ and correspondingly 
$f_G\sim1$. 
%considered 
%For the disk geometry  the accretion 
%occurs in a disk-like geometry, while it is close to$f_G\sim1 in case of  spherical accretion. 
%is assumed. 
%Despite this uncertainty 
% in the determination of $f_G$, 
%we adopt the above formula for  $f_G$ in which $\theta\sim i$ 
%if information on the system inclination angle $i$ is available (see Table \ref{tab:par_scal}).  

 %({\it top left} panel) 
In Figure~\ref{three_scal} we show   the $\Gamma-N_{bmc}$ correlation  for ESO~243--49 HLX--1 { (black 
points)} obtained  using  the $Swift$ %and $Suzaku$ 
spectra 
%BEGIN NEW
%[for two $N_H$ cases of the constant %$N_H^l=10^{21}$ cm$^{-2}$ 
%%($red$) and the variable %$N_H^h=5\times 10^{21}$ cm$^{-2}$ 
%`($black$)]
%}
%END NEW
 along with the correlations  for  { three Galactic reference sources  [XTE J1550-564 (blue), 4U~1630-47 (pink), and 
H~1743-322 (green), see left panel] and one extragalactic reference source M101 ULX-1 (red, see right panel)}.
%which  are similar  to the correlation found  for the target source.  
The BH masses and distances  for each of these target-reference pairs are shown in Table~\ref{tab:par_scal}. %7. 

The  BH mass, $M_t$  for HLX-1 can be evaluated using  a formula (see TS16)
\begin{equation}
M_t=C_0 {N_t} {d_t}^2 f_G %cos(\theta_r)/cos(\theta_ULX),
\label{C0 coefficient}
\end{equation}
\noindent where 
%the scale coefficient for each scaling pair    %coefficient 
$C_0=(1/d_r^2)(M_r/N_r)$ is the scaling coefficient for each of  the pairs (target and reference sources), 
masses $M_t$ and $M_r$ are in solar units and $d_r$ is the distance to a particular reference source  measured in kpc.

%{ Furthermore in Figure~\ref{three_scal} we plot the $\Gamma-N_{BMC}$ for M101 ULX-1 %187 
%points extracted using $Chandra$ and $Swift$ %and $Suzaku$ 
%spectra for two $N_H$ cases: the constant $N_H$ (red, in the {\it bottom} panels) and the variable $N_H$ ($black$, 
%in the {\it top right} panel) along with those for  the three reference patterns  [4U~1630-47 ($pink$), XTE J1550-564 ($blue$), 
%H~1743-322 ($green$)].  
%}
%which  are similar  to the correlation found  for our target source. 
%In Figure~\ref{three_scal} we also plot M101 ULX-1 data points 
% BEGIN NEW
%{$red$ and $blue$ for $N_H^l$ and $N_H^h$ respectively  
%}
% END NEW
%with their best-fit curves along 
%with those for the Galactic  BHs holes: %(target sources): 
%a) 4U~1630-47 (red triangles), 
% XTE~J1550-564 ($blue$ points in the $left$ panel) %(pink triangles) 
%and H~1743-322 ($black$ points in the $right$ panel). %(red stars). 
%In these diagrams, the target (M101 ULX-1) transition for scaling is shown in blue color. 
We used values of $M_r$, $M_t$, $d_r$, $d_t$, and $\cos (i)$ from Table~\ref{tab:par_scal} % 1 
and then  we calculated the lowest limit of the mass, using the best-fit value of  $N_t= (4.2\pm 0.1)\times 10^{-6}$ 
taken them at the beginning of the index saturation  (see Fig. \ref{three_scal}) and measured
%NEW
%{(for the constant $N_H$ case) and  $N_t=(2.5\pm 0.3)\times 10^{-5}$ 
%(for the variable $N_H$ case) 
in units of $L_{39}/d^2_{10}$ erg s$^{-1}$ kpc$^{-2}$ [see Table \ref{tab:parametrization_scal}
 for values of the parameters of function $F(N_t)$ (see Eq. 1)].
%}
%(see Fig. \ref{three_scal}). %$N_t=0.004\pm 0.001$ % (for $N_H^h$ case) and $N_t=0.006\pm 0.002$ (for $N_H^l$ case). 
Using $d_r$, $M_r$, $N_r$ (see ST09) we found that  $C_0\sim 2.0, 1.9, ~1.72$ and  $1.83$ 
%for XTE J1550, H 1743 and 4U 1630 respectively 
% of the three  reference sources using 
for M101 ULX-1, XTE J1550-564, H~1723-322 
{and 4U~1630-472}, respectively. 
% indicate on semi-consistency 
%of scaling approach.
%Then, using formula (\ref{C0 coefficient}),  
Finally,  we obtained that $M_{HLX}\ge 7.2\times 10^4~M_{\odot}$ 
($M_{HLX}=M_t$) %, {for $N_H^h$ case}) and 
%$M_{ULX}\ge 1.5\times 10^4~M_{\odot}$ (for $N_H^h$ case)}, 
assuming $d_{HLX}\sim$95 Mpc~\citep{Soria10} and  $f_G\sim1$.
% (inclinations for both objects are the same).  
We summarize all these results  in  Table~\ref{tab:par_scal}.
% BEGIN NEW
%To take account of the spread in the distance to M101, we have made the same estimates of $M_{ULX}$ assuming 
%$d_{ULX}=7.4\pm 0.6$ Mpc~\citep{Kelson96} and derived  higher values $M_{ULX}$
%$\ge 4.3\times 10^4~M_{\odot}$. %(for $N_H^h$ case) and $\ge 2\times 10^4~M_{\odot}$ (for $N_H^l$ case). 
%All these results %are obtained  for   the  constant  $N_H$ during the state transitions and they  
%are summarized in  Table~\ref{tab:par_scal}. %1 (see the lower part). 
%NEW
%{The  estimates for the case of the variable $N_H$ provide a lower values $M_{ULX}$: %. Namely, 
%$M_{ULX}\ge 0.7\times 10^4~M_{\odot}$  %($M_{ULX}=M_t$), %, {for $N_H^h$ case}) and 
%$M_{ULX}\ge 1.5\times 10^4~M_{\odot}$ (for $N_H^h$ case)}, 
%[assuming $d_{ULX}\sim 6.4$ Mpc~\citep{Shappee11}] and  % and if $f_G=1$ (inclinations for both objects are the same).  
%$\ge 0.9\times 10^4~M_{\odot}$ [for $d_{ULX}=7.4\pm 0.6$ Mpc~\citep{Kelson96}].
% BEGIN NEW
%To take account of the spread in the distance to M101, we made the same estimates of $M_{ULX}$ assuming 
%In turn, for the distance $d_{ULX}=7.4\pm 0.6$ Mpc~\citep{Kelson96} we derived  higher values $M_{ULX}$
%$\ge 0.9\times 10^3~M_{\odot}$. %(for $N_H^h$ case) and $\ge 2\times 10^4~M_{\odot}$ (for $N_H^l$ case).  
%}

% END NEW
It is worth noting that  the inclination of ESO~243-49 HLX-1 may be different from those  for the reference Galactic 
sources ($i\sim 60-70^{\circ}$), therefore we take this  BH mass estimate for ESO~243-49 HLX-1 as the  lowest BH mass 
value  because that $M_{HLX}$ is reciprocal function of $\cos (i_{HLX})$
% BEGIN NEW
[see Eq.~\ref{C0 coefficient} taking into account that $f_G=(\cos\theta)_r/(\cos\theta)_t$ there]. %In this context, we can suggest  that  the lower $N_H$ case corresponds to   
%the face-on system configuration, while the higher $N_H$ associated  with the higher  inclination  angle case.
% END NEW
%It is worth noting that if we use the low limit of absorption column $N_H=6\times10^{20}$ cm$^{-2}$ as suggested  by \cite{Liu09} we obtain the low limit of $M_{ULX}\gax6\times10^4$ solar masses.
%Finally, we present the estimated  values of  BH mass  %and the inclination angle 
%with the proper error bars  for the  %reference (XTE~J1550-564) and 
%target  source (M101 ULX-1)    
%are summarized 
% along with parameters for the reference  sources (XTE~J1550-564, H~1743-322, 4U~1630-47) in Table~\ref{tab:par_scal}. 
The obtained  BH mass estimate  agrees with a high bolometrical luminosity for ESO~243-49 HLX-1 and $kT_s$  value which is in the range of 50$-$140 eV. 
%low seed (disk) photon temperature of its X-ray spectrum. 
A very soft spectrum is consistent with  a relatively cold disk for a  compact object of high mass.
% ULXs that has also been considered 
%as  an evidence for IMBHs (Miller et al. 2003, 2004;  Wang et al. 2004). 
For example, Shakura \& Sunyaev, (1973) 
(see also Novikov \& Thorne, 1973) provide an effective temperature of the accretion material of 
$T_{eff}\propto M_{BH}^{-1/4}$.

Yan et al. (2015), suggested that the soft-to-hard state transition luminosity of HLX-1 is at 2\% of $L_{Edd}$ 
based on an analogy to the Galactic BHs, and estimated  the mass of the accreting compact object as $(8\pm 4)\times 10^4$ M$_{\odot}$. It is also important to emphasize  %nteresting 
that this original mass estimate for the central source in HLX-1, based on  a particular X-ray 
luminosity, is  consistent with our scaling BH mass estimate for HLX-1.
In addition, our HLX-1 mass estimate is also consistent with an  IMBH mass of $\sim 10^4 - 10^5$ M$_{\odot}$  derived 
using a detailed X-ray spectral modelling (Farrell et al. 2010; Davis et al. 2011; Servillat et al. 2011; 
Godet et al. 2012; Webb et al. 2012) 
and with the results obtained by Cseh et al. (2014) 
($M_{BH}<10^6$ M$_{\odot}$).

We  derived a bolometric luminosity  between 
$3\times 10^{41}$ erg/s and $4\times 10^{42}$ erg/s based on  the normalization of the BMC model.  
It is evident that this  high luminosity  is difficult to achieve in a X-ray binary unless the accretor has a mass greater than  10$^3$ $M_{\odot}$ to be consistent with the Eddington limit. %While 
%it is possible that 
Our luminosity estimate  agrees with that %higher than that for 
previously obtained for HLX-1 with different instruments  (see Farrell et al. 2009; Godet et al. 2009; Yan et al. 2015).
However,  using optical observations (FORS2 spectrograph
on the Very Large Telescope),  Soria et al. (2013) found that the H$_{\alpha}$ emission from HLX-1 ($L_{H_{\alpha}}\approx$ a few times of $10^{37}$ erg/s) 
could be excited by  X-ray luminosity of $\sim  10^{40}$ erg/s, which
is an order of magnitude smaller than the mean luminosity observed over 2009 -- 2012.
 Soria et al. suggested  that the observed $H_{\alpha}$ emission comes not from the disk surface (it does not have a disk-like profile) but from some material further out, or perhaps
from the remnants of previous outflows.

\subsection{Possible  effects on an estimate of BH mass in ESO~243--49   \label{caveats}}

Now we discuss some potential sources of systematic errors which can affect the validity
of our method.

\subsubsection{Suggestion on  a similarity of 
%the main physical processes in  
Galactic BHs binaries and HLX$-$1 
\label{caveats_similarity}}

This similarity 
%of main physical processes for both Galactic BH binaries and HLX--1 
is based on comparative analysis of spectral properties for these two classes of objects that are revealed  during X-ray outbursts
% in these sources 
(see Sect.~\ref{HID_lc}). In particular, we  clearly observe  softening of X-ray emission 
with X-ray outburst 
flux in HLX--1 (e.g., Fig.~\ref{HID}), which is similar to that in the bright Galactic low-mass X-ray binaries.

The hardness-intensity  diagram shown in  Fig.~\ref{HID} are model independent. 
We demonstrate that  spectra of HLX--1 are poorly fitted by {\it power-law, bbody} models and their combination (see Table~\ref{tab:par_swift}). The detailed modeling of X-ray spectral shape reveals  a strong rise of the low-energy component 
(for photon energies less than  1 keV) along with  a steepening  of the higher energy tail during outburst development. This observational effect  is a strong confirmation of the converging inflow paradigm in the case of a BH source [see \cite{st09}].

\subsubsection{Validity of BH mass determination in ESO~243--49 using Galactic BH mass values \label{caveats_gal_BH_mass}}

In this paper, we used Galactic X-ray binaries to compare their spectral characteristic with those established in ESO 243--49 HLX--1.  We used  XTE J1550--564  for which the mass of compact object 
(BH) is evaluated applying  ``gold standard'' dynamical measurements (Orosz, 2002).
Therefore,  BH mass scaling for a pair of ESO 
243--49 HLX--1 and XTE J1550--564 provides a reliable BH mass estimate for  ESO 243--49 HLX--1. However, the BH  mass values of other objects used for scaling procedure, such as 4U~1630--47, H~1743--322, are not based on dynamical 
measurements and can not be considered as ``traditional'' %solid 
BH estimates. 
%While they rather used for additional 
%illustration of what how scaling method works, 
It is worth noting that the BH mass value  for ESO~243--49  using  these BH estimates for 4U~1630--47, H~1743--322 well  agrees  
with that  applying  BH mass of  XTE J1550--564.

\subsubsection{Validity of BH mass estimate in ESO~243--49 using scaling with M101 ULX--1 \label{caveats_M101_ULX-1_mass}}

We used the extragalactic source M101 ULX--1 for a BH mass estimate in ESO 243--49. 
 We should point out a wide range of the BH mass estimates 
for M101 ULX--1 obtained by applying  X-ray (Kong et al. 2004; Kong \& Di Stefano, 2005; TS16) and optical data (Liu et al. 2013).   In particular, Kong et al. (2004) based on $Chandra$ and XMM-$Newton$ observations of this source 
%M101 ULX--1 
during the 2004 July outburst, obtained an estimate for the BH mass greater than 2800 M$_{\odot}$, while, Kong \& 
Di Stefano (2005) suggested  that  $M_{m101}$ is in the range of $1.3\times 10^3 - 3\times10^4$ M$_{\odot}$. In our recent paper (see TS16)  we found, 
using $Swift$ (2006 -- 2013) and $Chandra$ (2000, 2004, 2005) data, that the BH mass in this source is 
%  also indicated on high BH mass 
of order of $\sim 10^4$ 
M$_{\odot}$ using scaling method (ST09) (which is comparable with the Kong \&  Di Stefano's BH values). 
On the other hand, based on HST optical data of M101 ULX--1 %with HST 
and using the dynamical method, Liu et al. (2013) estimated  mass of the compact object as 5 -- 1000 %20 -- 30 
M$_{\odot}$.
% depending on   %suggesting high 
%an inclination angle $i$ (for an interval from 80 to 9 degrees, correspondingly).  
We note that Liu et al. (2013) made this BH mass estimate based on radial velocity analysis and adopted a simple two-point mass model for a binary  without taking into account a tidal influence and a heating effect to  the optical star by the X-ray companion, which can significantly change  the resulting BH mass value \citep{Ant_Cher94,Ant16,Petrov16}.

This shows  a tendency for smaller values of BH mass, $M^{opt}_{m101}$ using optical data and  higher values $M^{X}_{m101}$    when  X-ray data are used. 
%for M101 UX--1. 
We  estimated the BH mass in ESO 243--49, $3\times 10^3< M_{ESO} < 7\times10^4$  M$_{\odot}$ by 
applying the scaling technique with X-ray mass 
$M^{X}_{m101}$ ($2.8\times 10^3 - 3\times 10^4$ M$_{\odot}$, see TS16).  
% and see Table~\ref{tab:par_scal} in this paper), 
%we obtain BH mass within an interval of . 
When we  applied  an optical BH mass  estimate  
$M^{opt}_{m101}$ ($5 - 10^3$ M$_{\odot}$), we  found    lower masses within a wide interval  of 
from 7 M$_{\odot}$ to  $2\times 10^3$ M$_{\odot}$  for  $M_{ESO}$ which is inconsistent 
with so called  fundamental plane' 
results $10^3< M_{ESO} < 10^5$ M$_{\odot}$ (see Davis et al. 2011 and  Godet et al. 2011).
% which is only barely consistent with ``fundamental  plane'' 
%results $10^3< M_{ESO} < 10^5$ M$_{\odot}$. 
we therefore conclude that our scaling method  applied to the pair of BH sources ESO 243--49 and M101 ULX--1  leads to  better constraints  with  the case of  the X-ray BH mass %$M^{X}_{m101}$ 
than that  of optical BH mass.

We  note that the massive BHs  in   ESO 243--49 and M101 ULX--1   are not located   in galactic nuclei as  it should be in the case of supermassive BHs (SMBHs). 
As  is known, the stellar bulge of almost every massive galaxy
contains a SMBH (Ferrarese \& Ford 2000). When  galaxies merge  they can form two or more SMBHs with  their stellar bulges  which are  outside of the galactic centre.
%offset in position with the respect . 
 %stellar bulges initially containing two or more SMBHs that are offset
%in position. 
Following a merger, a pair of inspiraling SMBHs can remain   for a while in  separation
 before forming a tight binary or 
%that could either stall at some separation or
finally merge (see Begelman et al. 1980). Similarly,  the high  mass of some IMBHs can be formed as a result of   merging of  galaxy nuclei. This scenario  can be a valid argument 
for large masses in  M101 ULX--1 and ESO 243--49.

 There are still many open quiestions from a theoretical point of view how
to explain a formation of   massive BHs (such as IMBHs). Recently, Latif \& Ferrara (2016) discussed possible
formation mechanisms of supermassive BHs. In particular, they suggested that
``seed'' BHs were  formed early on, and grow  either through   rapid accretion or
BH/galaxy mergers.
% taking into account effects of metallicity and rotation. 
Latif \& Ferrara  offered  three most popular BH formation scenarios:
dynamical evolution of dense nuclear star clusters, a core-collapse of massive
stars, and a collapse of a protogalactic metal free gas cloud.

We included M101 ULX--1 in a scaling BH mass sample for ESO 243--49, 
%in spite of the fact that 
even though  Soria \& Kong (2016) 
presented strong arguments that the observed emission in M101 ULX--1 is not a classic accretion disk.  Specifically, Soria \& Kong  re-examined the X-ray spectral and timing properties of M101 ULX--1 using  a series of $Chandra$ and 
XMM-$Newton$ observations and showed that their  model of 
%the photosphere of 
an optically thick outflow is consistent  with the data. 
They showed that the characteristic radius, $R_{BB}$ of a thermal emitter and its color 
temperature, $kT_{BB}$  are  approximately related to each other as $R_{BB}\propto T^{-2}_{BB}$. In addition, they revealed absorption edges 
fitting the M101 ULX--1 spectra  applying thermal plasma models.  They interpreted  this modeling along with  the data  as an evidence of a clumpy, 
multi-temperature outflow around ULX--1, in  particularly in the HSS.  Soria \& Kong highlighted that M101 ULX--1 belongs to ultraluminous supersoft sources (ULSs) rather than to 
%ultraluminous X-ray source 
ULX population. It might therefore be argued that if  Soria \& Kong are correct then the method of the BH mass determination applied here might not be applicable for M101 ULX--1.

Our arguments which are
%and BH mass estimates 
based on 
 %find 
the correlation of the  photon index with  the mass accretion rate   in M101 ULX--1 and its resemblance  with those  in a number of Galactic sources allow us argue that the innermost part of the accretion flow (the disk-Compton cloud-converging flow configuration)  is similar in these sources  (see TS16 and Fig. \ref{geometry} here). Moreover,  this index-mass accretion rate correlation for M101 ULX-1 is almost identical in terms of the shape for that in ESO 243--49 HLX--1
% then one can see that this correlation will not be similar with those ESO 243--49 HLX--1 
and those in the chosen Galactic BHs. It is not by chance that 
we compared all these correlations with each other and found that they are self-similar.

It is worthwhile to point out that \cite{soria16}   described the  hard tail   of the 
{\it Chandra}  ULX spectrum of  M101 ULX-1 using higher temperatures. 
%were recently attempted by \citep{soria16} in application to M101 ULX-1 observed
%with $Chandra$. In this way, 
%Soria \& Kong (2016) re-examined $Chandra$ and XMM-$Newton$ data for super-soft
%ULX-1 in M101 and 
They revealed  a relatively hard tail in the 0.3 -- 6 keV energy range and 
%% source spectra. Specifically, to fit high
%%count-rate spectra of M101 ULX-1 
 applied three {\it mekal} components in addition to sample a soft {blackbody} component.
The high count-rate spectra of  three of these {\it mekal}
components are associated with  temperatures of 0.6 keV, 1 keV and $\ge$2 keV. 
However, when Soria \& Kong replaced this multiple component model 
%the three temperature mekal 
%components 
by the single $Comptt$~\citep{t94} component (see  Table A3 there),
they {also} obtained a good quality fit with the seed photon temperatures in the range of
90 -- 135 eV. 

\section{Conclusions \label{summary}} 

We found the low$-$high state transitions observed in  HLX-1 using the full set of $Swift$-/XRT  
observations (2008 -- 2015) and %, %observations during  2006 -- 2013, 
%$Suzaku$ (2006) and $Chandra$ (2000, 2004 -- 2005) observations. 
we showed  the observed spectra can be fitted  by the BMC model   
%that the source spectra can be fitted by 
%the BMC model 
for all observations, independently of the spectral state of the source. 
%NEW
%{We have also investigated a possible absorption effect in X-ray spectra of M101 ULX-1 
%(by neutral matter) taking into account two possible cases for the absorbing column: 
%($i$) the constant  $N_H$ %along state transition 
%and 
%($ii$) the variable $N_H$ %ranging from 0.1 to $5\times 10^{21}$ cm$^{-2}$ 
%during the state transition. 
%}

We investigated the X-ray outburst properties of HLX-1 and confirmed the presence of spectral state transitions 
during the outbursts using of hardness-intensity diagrams (Godet et al. 2009; Servillat et al. 2011) and 
the index$-$normalization (or $\dot M$) correlation observed in HLX-1, which were similar to those in Galactic BHs. 
In particular, we found that HLX-1 follows the $\Gamma-\dot M$ correlation previously obtained for extragalactic IMBH source 
M101 ULX-1 and Galactic BHs,  4U~1630-472, XTE~J1550-564 and H~1743-322  taking into account the %different 
particular values of the $M_{BH}/d^2$ ratio (Figure~\ref{three_scal}).
%(Godet et al. 2009; Servillat et al. 2011), similar to the cycle of low/hard and high/soft states in Galactic XRBs. 
%Our study reveals that the index$-$normalization (or $\dot M$) correlation observed in HLX-1 is similar to those in GBHs. 
%sources, also holds
%for : variations of 
The photon index $\Gamma$ of ESO~243-49 HLX-1 spectrum is  in the range $\Gamma = 1.6 - 3.0$. 
We  also estimated the peak bolometric luminosity, which is about $4\times 10^{42}$ erg s$^{-1}$.  
%BEGIN NEW
%{while it is  lower, and about $1.5\times 10^{41}$ erg s$^{-1}$ for the lower $N_H$ case ($N_H= 10^{21}$ cm$^{-2}$).
%}  
% END NEW

%We applied the scaling technique based on the observed correlations 
 We used the observed index-mass accretion rate correlation to estimate $M_{BH}$  in HLX-1. 
This scaling method was  successfully implemented  to find  BH masses of Galactic (e.g. ST09, STS13) 
and extragalactic black holes [TS16; \citet{sp09}; \cite{ggt14}].  
An application of the scaling  technique  to the X-ray data  from XRT/Swift observations of  ESO~243-49 HLX-1 allowed us 
%In this work the scaling technique for  the first time  is applied  
to estimate  $M_{BH}$ for this particular source. 
We found  values of 
%BEGIN NEW
$M_{BH}\geq 7.2\times 10^4 M_{\odot}$.  

Furthermore, our  BH mass estimate agrees %close to 
 the previously established IMBH mass of $\sim 10^4 - 10^5$ M$_{\odot}$ derived using the detailed X-ray spectral modelling (Farrell et al. 2010; Davis et al. 2011; Servillat et al. 2011; Godet et al. 2012; Webb et al. 2012).

Combining all these estimates with  the inferred  low temperatures of the seed disk photons $kT_s$, we can state  that 
%in X-ray spectrum of M101 ULX-1, 
% that M101 ULX-1 is close to star forming region and 
%has an optical counterpart consistent with a B supergiant/WR star, as well as scaling mass estimates, 
the compact object of  ESO~243-49 HLX-1 probably is  an intermediate-mass black hole with a mass at least $M_{BH}> 7.2\times10^4 M_{\odot}$.

This research was performed using  data supplied by the UK $Swift$ Science Data Centre at the University of Leicester.  We also acknowledge the interesting remarks and points of the referee.

\newpage

%
%Table 1
%

\begin{deluxetable}{l l l l l c}
%%%%%\rotate
\tablewidth{0in}
\tabletypesize{\scriptsize}
%  \begin{center}
    \tablecaption{
List of $Swift$ observations of ESO 243-49 HLX-1}
    \renewcommand{\arraystretch}{1.2}
%    \begin{tabular}[h]
%      \hline
\tablehead{
 Obs. ID& Start time (UT)  & End time (UT) &MJD interval}
%Satellite&Obs. ID& Start time (UT)  & End time (UT)}
%%%%%Obs.  &ID           & time (UT)& time (UT)& of state& }
\startdata
 00031287(001-164, 233, 235-238 & 2008 Oct. 24 & 2012 Sept.  &54763.6$-$56198.9& \\
240-246, 248-250, 252, 254, 255)$^{1,2,3,4}$  &  &  &         & \\
00032577(001-100)$^4$  & 2012 Oct. 2  &2015 Apr. 10 & 56202.5 -- 57122.5 & \\
00049794(001, 003)$^4$ & 2013 March 14 & 2013 March 17  &56365.3 -- 56369.3 &\\
00080013001$^4$ & 2012 Nov. 21 02:05:59 & 2012 Nov. 21 07:12:03  & 56252.1 -- 56252.3 & \\
00091907(006-037)$^4$ & 2014 May 15 & 2015 March 5 & 56792.1 -- 57086.5 & \\
00092116(001-020) & 2015 April 5  & 2015 Sept. 22  & 57117.7 -- 57287.1 & \\
%20889003& 1999 Aug. 19 02:01:32 & 1999 Aug. 20 04:54:32 &51409.1-51410.2$^1$& \cite{piraino00}\\
      \enddata
%      \hline
%      \end{tabular}
   \label{tab:table}
% \end{center}
{\bf References}, 
(1) Soria et al. 2010; 
(2) Farrell et al. 2013;
(3) Webb et al. 2010, 2014;
(4) Yan et al. 2015.
%(3) ??? 
%(1) \cite{piraino00} 
%Piraino et al., (2000)
\end{deluxetable}

\newpage

%
%Table 2
%

\begin{deluxetable}{lllllllll}
%\rotate
\tablewidth{0in}
\tabletypesize{\scriptsize}
%  \begin{center}
    \tablecaption{ 
Best-fit parameters  of the combined $Swift$ spectra
 of ESO 243-49 HLX-1 in the 0.3$-$10~keV  range using the following four %two 
models$^\dagger$: phabs*power-law, phabs*bbody, phabs*(bbody+power-law) and phabs*bmc 
}
    \renewcommand{\arraystretch}{1.2}
%    \begin{tabular}[h]
   %\hline
%\hline
%& Parameter & Band-A & Band-B & Band-C & Band-D & Band-E & Band-F & Band-G \\
\tablehead{Parameter & Band-A & Band-B & Band-C & Band-D & Band-E & Band-F & Band-G}
%\hline 
 \startdata
Hardness ratio  &   HR        & $>1$        & $0.5-1$                &   $0.25-0.5$& $0.13-0.25$  &     $0.07-0.13$ & $0.03-0.07$ & $<0.03$ \\
 \hline                                       %inserts single line
Model &      &        &        &       &      &      &     &   \\
\hline                                             %inserts single line
phabs       & N$_H$             & 5.1$\pm$0.1 & 5.2$\pm$0.2 & 5.1$\pm$0.1  & 5.1$\pm$0.1 & 5.2$\pm$0.1 & 5.1$\pm$0.1  & 5.05$\pm$0.08 \\
Power-law   & $\Gamma_{pow}$    & 1.4$\pm$0.1 & 1.6$\pm$0.2 & 1.9$\pm$0.2  & 2.4$\pm$0.1 & 2.6$\pm$0.2 & 2.8$\pm$0.3 & 3.9$\pm$0.4 \\
          & N$_{pow}^{\dagger\dagger}$ & 0.24$\pm$0.05 & 0.43$\pm$0.02 & 0.97$\pm$0.05 & 2.7$\pm$0.4 & 3.04$\pm$0.06 & 4.3$\pm$0.1 & 6.2$\pm$0.1 \\
          & $\chi^2$ {\footnotesize (d.o.f.)} & 1.15 (138)     & 1.1 (180)     & 1.3 (209) & 1.5 (223)   & 2.6 (250) & 3.2 (93)     & 2.1 (219)    \\% & 1.8 (19)     & 1.1 (19)  \\
     \hline
phabs      & N$_H$             & 5.0$\pm$0.1 & 5.0$\pm$0.2 & 4.9$\pm$0.2  & 5.0$\pm$0.1 & 5.0$\pm$0.09 & 4.9$\pm$0.2  & 5.03$\pm$0.06 \\
%& 0.03$\pm$0.01 & 0.05$\pm$0.01 & 0.03$\pm$0.01 \\ 
bbody      & T$_{BB}$          & 280$\pm$10   & 170$\pm$5  & 85$\pm$9    & 97$\pm$6  & 110$\pm$9         & 85$\pm$7   & 120$\pm$10 \\
           & N$_{BB}^{\dagger\dagger}$ & 0.53$\pm$0.04  & 0.9$\pm$0.3   & 2.1$\pm$0.5   & 3.1$\pm$0.5 & 3.5$\pm$0.4   & 4.1$\pm$0.5 & 5.0$\pm$0.3  \\
           & $\chi^2$ {\footnotesize (d.o.f.)} & 6.1 (138)      & 4.5 (180)    & 3.8 (209)  & 2.56 (223)   & 1.4 (250) & 1.2 (93)     & 1.1 (219)    \\
\hline
phabs      & N$_H$             & 5.1$\pm$0.1 & 5.1$\pm$0.1 & 5.1$\pm$0.08  & 5.1$\pm$0.1 & 5.2$\pm$0.2 & 5.1$\pm$0.1  & 5.08$\pm$0.09 \\
bbody      & T$_{BB}$          & 300$\pm$10   & 180$\pm$9  & 90$\pm$6    & 100$\pm$20  & 120$\pm$8         & 92$\pm$5   & 110$\pm$8 \\
          & N$_{BB}^{\dagger\dagger}$ & 0.34$\pm$0.05  & 0.7$\pm$0.2   & 1.8$\pm$0.6   & 2.6$\pm$0.4 & 4.3$\pm$0.5   & 4.9$\pm$0.6 & 5.0$\pm$0.2  \\
Power-law  & $\Gamma_{pow}$    & 1.3$\pm$0.2 & 1.7$\pm$0.3 & 1.8$\pm$0.1   & 2.3$\pm$0.2 & 2.4$\pm$0.2 & 2.6$\pm$0.1 & 2.9$\pm$0.3 \\
           & N$_{pow}^{\dagger\dagger}$ & 0.82$\pm$0.03& 0.65$\pm$0.03 & 0.48$\pm$0.09    & 0.39$\pm$0.07 & 0.4$\pm$0.4 & 0.67$\pm$0.02& 0.52$\pm$0.03 \\
          & $\chi^2$ {\footnotesize (d.o.f.)}& 1.24 (136) & 1.18 (178) & 1.24 (207)& 1.23 (221)& 1.26 (248) & 1.19 (91) & 1.14 (217) \\
\hline
phabs      & N$_H$             & 5.0$\pm$0.1 & 5.0$\pm$0.1 & 4.9$\pm$0.1  & 5.0$\pm$0.1 & 5.0$\pm$0.1 & 4.9$\pm$0.1  & 5.02$\pm$0.04 \\
bmc        & $\Gamma_{bmc}$    & 1.6$\pm$0.2 & 1.76$\pm$0.09& 2.01$\pm$0.05   & 2.68$\pm$0.08    & 2.8$\pm$0.1    & 2.96$\pm$0.09    & 3.0$\pm$0.1 \\
          & T$_{s}$           & 54$\pm$9       & 61$\pm$8    & 52$\pm$9       & 139$\pm$8       & 142$\pm$10    & 105$\pm$9      & 130$\pm$10   \\
          & logA$$            & 0.10$\pm$0.04  & 0.17$\pm$0.05& 1.1$\pm$0.3    & -1.09$\pm$0.05  & -1.22$\pm$0.09& -1.06$\pm$0.05   & -0.7$\pm$0.3 \\
          & N$_{bmc}^{\dagger\dagger}$ & 0.36$\pm$0.09 & 0.6$\pm$0.2 & 0.98$\pm$0.07  & 2.5$\pm$0.3  & 3.08$\pm$0.06 & 4.03$\pm$0.05& 5.2$\pm$0.1 \\
           & $\chi^2$ {\footnotesize (d.o.f.)}& 0.86 (136) & 0.89 (178) & 0.96 (207)& 0.93 (221)& 1.03 (248) & 0.79 (91) & 1.14 (217) \\
%      \hline
 %\hline  
\enddata
%      \hline
%      \end{tabular}
    \label{tab:par_swift}
%  \end{center}
$^\dagger$     Errors are given at the 90\% confidence level. %this parameter is fixed, 
$^{\dagger\dagger}$ The normalization parameters of blackbody and bmc components are in units of $L^{soft}_{33}/d^2_{10}$ erg s$^{-1}$ kpc$^{-2}$, 
where $L^{soft}_{33}$ is the soft photon luminosity in units of $10^{33}$ erg s$^{-1}$, $d_{10}$ is the distance to the 
source in units of 10 kpc, and oower-law component is in units of 10$^{-6}$ %photons 
keV$^{-1}$ cm$^{-2}$ s$^{-1}$ at 1 keV. 
%$^{\dagger\dagger\dagger}$ spectral flux 
%in the 3-- 150 energy range 
%in units of $\times 10^{-15}$ erg/s/cm$^2$. 
$N_H$ is the column density for the neutral absorber, {in units of $10^{20}$ cm$^{-2}$} (see details in the text).
% $5\times 10^{20}$ cm$^{-2}$ (see details in the text). %   in units of $10^{22}$ cm$^{-2}$. %($\times 10^{22}$ cm$^{-2}$).
$T_{BB}$ and $T_{s}$ are the temperatures of 
the blackbody and seed photon components, respectively (in eV). $\Gamma_{pow}$ and $\Gamma_{bmc}$ are the indices of the {\it power law} 
and bmc, respectively.
\end{deluxetable}

\newpage
\begin{deluxetable}{lcccccc}
%%%%%\rotate
\tablewidth{0in}
\tabletypesize{\scriptsize}
%  \begin{center}
    \tablecaption{Parameterizations for the reference and target sources}
    \renewcommand{\arraystretch}{1.2}
%    \begin{tabular}[h]
%      \hline
\tablehead{
 Reference source & $\cal A$ & $\cal B$ & $\cal D$ & $x_{tr}$   & $\beta$ }
%Satellite&Obs. ID& Start time (UT)  & End time (UT)}
%%%%%Obs.  &ID           & time (UT)& time (UT)& of state& }
\startdata
XTE~J1550-564 RISE 1998 & 2.84$\pm$0.08 &  1.8$\pm$0.3    &  1.0 & 0.132$\pm$0.004   &   0.61$\pm$0.02  \\
H~1743-322    RISE 2003 & 2.97$\pm$0.07 &  1.27$\pm$0.08  &  1.0 & 0.053$\pm$0.001   &   0.62$\pm$0.04  \\
4U~1630-472   & 2.88$\pm$0.06 &  1.29$\pm$0.07  &  1.0 & 0.045$\pm$0.002   &   0.64$\pm$0.03  \\
M101 ULX-1   & 2.88$\pm$0.06 &  1.29$\pm$0.07   & 1.0  &   [4.2$\pm$0.2]$\times 10^{-4}$ &   0.61$\pm$0.03  \\%& $constant$ \\
 \hline
\hline  
 Target source     &      $\cal A$     &    $\cal B$    &  $\cal  D$  &   $x_{tr} [\times 10^{-6}]$ & $\beta$ \\%& $N_H$ hypothesis \\
  \hline
ESO 243-49 HLX-1 & 3.00$\pm$0.04 & 1.27$\pm$0.05   & 1.0  &   4.25$\pm$0.03 & 0.62$\pm$0.05  \\
 \enddata
%      \hline
%      \end{tabular}
   \label{tab:parametrization_scal}
\end{deluxetable}

\newpage
% Table 4

\begin{deluxetable}{lllllc}
%%%%%\rotate
\tablewidth{0in}
\tabletypesize{\scriptsize}
%  \begin{center}
    \tablecaption{BH masses and distances}
    \renewcommand{\arraystretch}{1.2}
%    \begin{tabular}[h]
%      \hline
\tablehead{Source   & M$^a_{dyn}$ (M$_{\odot})$ & i$_{orb}^a$ (deg) & d$^b$ (kpc)  & $M_{lum}$ (M$_{\odot}$) &M$_{scal}$ (M$_{\odot}$) }
 %Reference source & $\cal A$ & $\cal B$ & $\cal D$ & $x_{tr}$   & $\beta$ }
%Satellite&Obs. ID& Start time (UT)  & End time (UT)}
%%%%%Obs.  &ID           & time (UT)& time (UT)& of state& }
\startdata
XTE~J1550-564$^{(1, 2, 3)}$  &   9.5$\pm$1.1 &  72$\pm$5    &   $\sim$6           &...&   10.7$\pm$1.5$^c$ \\
H~1743-322$^{(4)}$     &   ...    &  75$\pm$3    &   8.5$\pm$0.8          &...&   13.3$\pm$3.2$^c$ \\
4U~1630--47$^{(4+1)}$    &       ...     &   $\leq$70   &   $\sim$10 -- 11    &...&   9.5$\pm$1.1     \\
%& $constant$ \\
M101~ULX-1$^{(6, 7, 8, 9, 10)}$     & 5 -- 1000     &     ...      & (6.4$\pm$0.5)$\times 10^3$, (7.4$\pm$0.6)$\times 10^3$ &...& $\ge 3.2\times10^{4}$, $\ge 4.3\times10^{4}$ \\
ESO 243-49 HLX-1$^{(11, 12)}$  & ... &     ...      & $\sim$95$\times 10^3$ &8$\pm 4\times 10^4$& $\ge 7.2\times10^{4}$ \\
 \hline
  \enddata
%      \hline
%      \end{tabular}
   \label{tab:par_scal}
{\bf References}. 
(1) Orosz et al. 2002; 
(2) S$\grave a$nchez-Fern$\grave a$ndez et al. 1999; 
(3) Sobczak et al. 1999;  
(4) Petri 2008; 
(5) STS14;
(6) Shappee \& Stanek 2011;
(7) Mukai et al. 2005; 
(8) Kelson et al. 1996;
(9) TS15;
(10) Liu et al. (2013); 
(11) Farrell et al, 2009;
(12) Soria et al. 2013
\\
{\bf Notes}.$^a$ Dynamically determined BH mass and system inclination angle, $^b$ Source distance found in literature, 
$^c$ Scaling value found by ST09.
\end{deluxetable}

%  FIgure 1
%

\newpage 

\newpage 

\newpage 

\newpage 
\begin{figure}[ptbptbptb]
  \includegraphics[width=17cm]{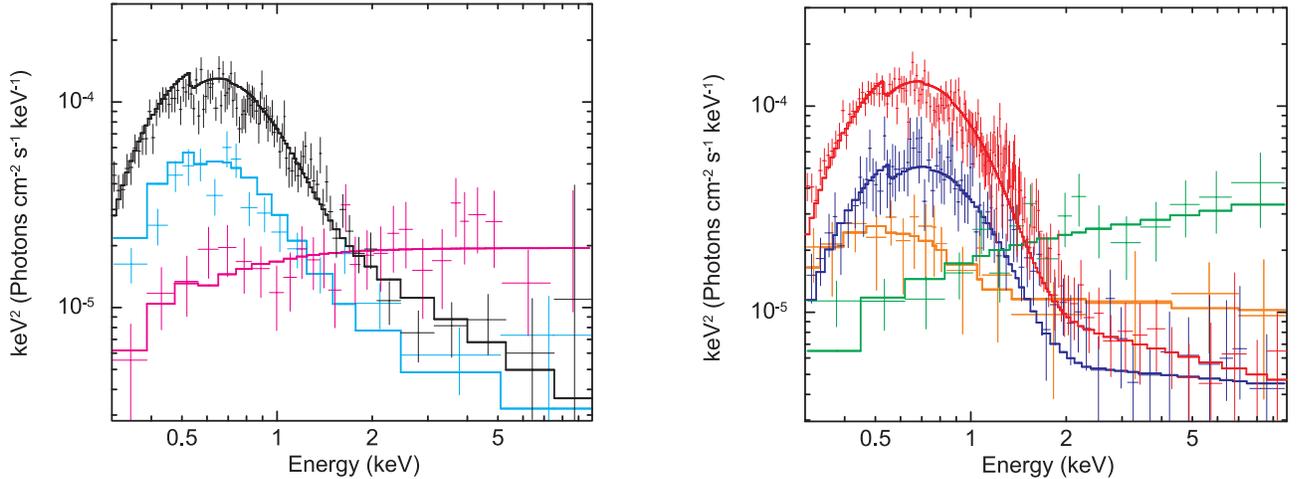}
      \caption{
Seven %Six %representative 
$EF_E$ spectral diagrams which are related to different spectral states  
of ESO 243-49 HLX-1 using the BMC model. The  data are taken from XRT/$Swift$  
observations related to  different {hardness ratios}: 
(left:) HR<0.03 (black, HSS), 0.03<HR<0.07 ({bright blue}, HSS), 0.5<HR<1 (pink, LHS); 
(right:) 0.1<HR<0.2 (blue, IS), 0.07<HR<0.2 (red, HSS), 0.2<HR<0.5 (orange, IS), HR>1 (green, LHS).
}
\label{6_swift_sp_compar}
\end{figure}

\newpage 
\begin{figure}[ptbptbptb]
\includegraphics[scale=0.75, angle=0]{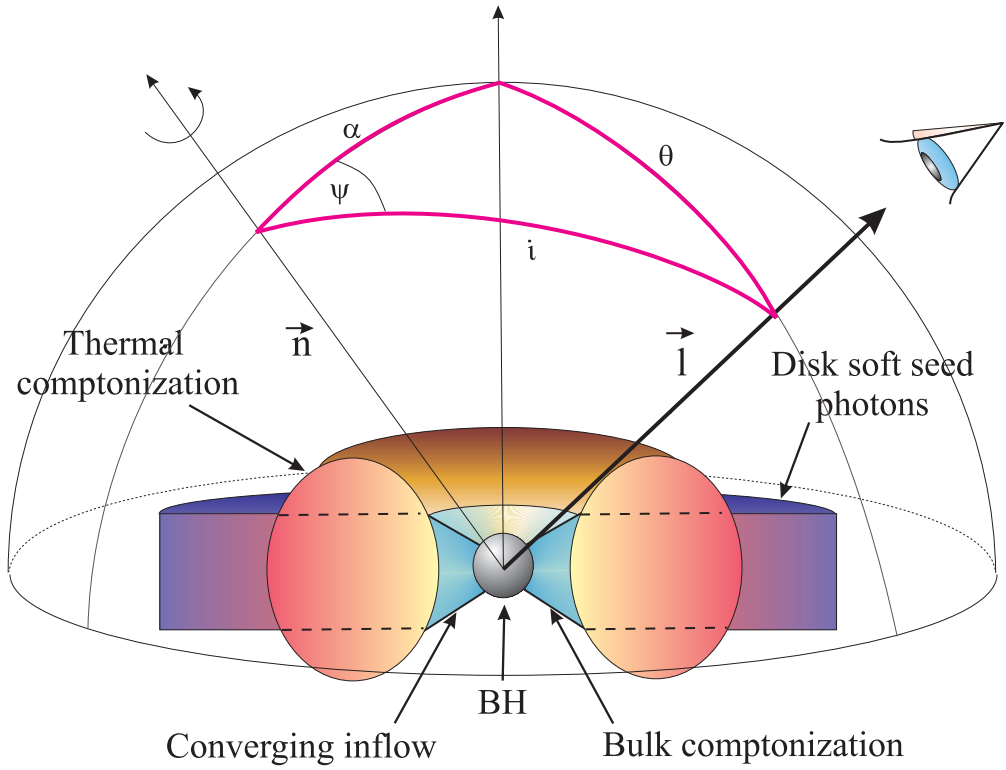}
\caption{A suggested  geometry of the system.   Disk  
%neutron star 
soft photons are upscattered (Comptonized) off  relatively hot plasma of the transition layer.  %Some fraction of these photons are  directly seen  by the Earth observer.  Blue and pink  photon trajectories correspond to soft (disk) and hard (Comptonized)  photons respectively.
}
\label{geometry}
\end{figure}

\newpage 
%\begin{figure}
%  \begin{figure*}
% \centering
%    \includegraphics[width=17cm]{f5_eso.eps}
%      \caption{Examples of %typical 
%E*F(E) spectral diagram of ESO 243--49 HLX--1 during the $hard$, $intermediate$ and $soft$ %($blue$) and  %($red$) 
%state events. The best-fit $Swift$ spectra (top panel) using the $BMC$ model, % wabs ? (Blackbody + CompTB + Gaussian) 
%along with $\Delta\chi$ (bottom panel) for the %very 
%hard %low luminosity 
%(band-B) state ($\chi^2_{red}=0.89$ for 178 % 136 
%d.o.f.), for the intermediate (band-E) state ($\chi^2_{red}=1.03$ for 248 d.o.f.) %$red$) 
%and for the soft %the very high luminosity 
%(band-G) state ($\chi^2_{red}=1.04$ for 317 d.o.f.). %, $blue$). 
%The best-fit model
%parameters are $\Gamma=1.76\pm 0.09$, $T_s=61\pm 8$ eV  %, $N_{bmc}=0.36\pm0.09$ and $logA$
% (for the low hard state),  $\Gamma=2.8\pm 0.1$, $T_s=142\pm 10$ eV (for the intermediate state,  
%and $\Gamma=3.0\pm 0.1$, $T_s=130\pm 10$ eV (for the high soft state,  
%see more details in Table~\ref{tab:par_swift}).
%}
%\label{two_state_spectra}
%\label{2_spectra}
%\end{figure}
%\end{figure*}

\begin{figure}[ptbptbptb]
\includegraphics[scale=0.95, angle=0]{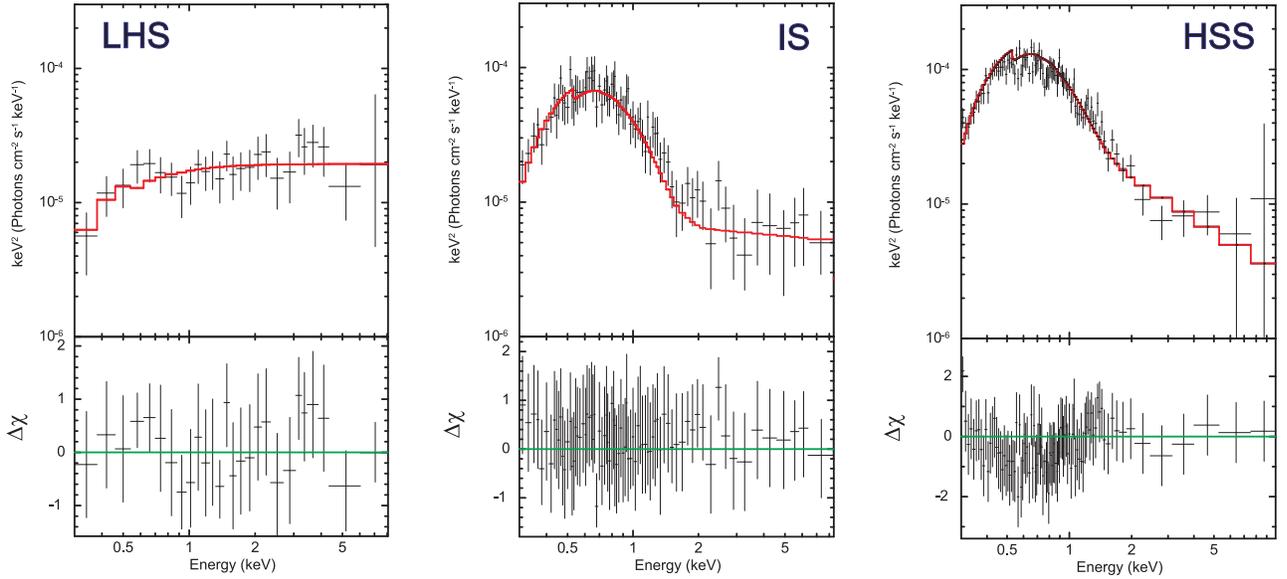}
\caption{Examples of %typical 
E*F(E) spectral diagram of ESO 243--49 HLX--1 during the hard, intermediate, and soft %($blue$) and  %($red$) 
state events. The best-fit $Swift$ spectra (top panel) using the BMC model, % wabs ? (Blackbody + CompTB + Gaussian) 
along with $\Delta\chi$ (bottom panel) for the %very 
hard %low luminosity 
(band-B) state ($\chi^2_{red}=0.89$ for 178 % 136 
d.o.f.), for the intermediate (band-E) state ($\chi^2_{red}=1.03$ for 248 d.o.f.) %$red$) 
and for the soft %the very high luminosity 
(band-G) state ($\chi^2_{red}=1.04$ for 317 d.o.f.). 
The best-fit model
parameters are $\Gamma=1.76\pm 0.09$, $T_s=61\pm 8$ eV  %, $N_{bmc}=0.36\pm0.09$ and $logA$
 (for the low/hard state),  $\Gamma=2.8\pm 0.1$, $T_s=142\pm 10$ eV (for the intermediate state,  
and $\Gamma=3.0\pm 0.1$, $T_s=130\pm 10$ eV (for the high/soft state,  
see more details in Table~\ref{tab:par_swift}).
%Evolution of ASM/{\it RXTE} count rate %, flux density $S_{8.46 GHz}$ at 8.46 GHz (VLA), 
%BMC normalization and 
%photon index $\Gamma$ 
%during  1996 -- 2011 observations of 4U~1630--47. 
%{\it Blue} vertical strips (at {\it top of the panel}) indicate temporal distribution of the {\it RXTE} data 
%of pointed observations used in our analysis, whereas {\it bright blue} rectangles indicate
% (trace) 
%the {\it RXTE} data sets listed in Table 2, 
%and {\it green} triangles show {\it Beppo}SAX NFI data, listed in Table 1.
 }
\label{two_state_spectra}
\end{figure}

%                                                                                             
%  FIgure 2  
%

% 
%  FIgure 3
%

\newpage 
\begin{figure}[ptbptbptb]
\includegraphics[scale=0.95,angle=0]{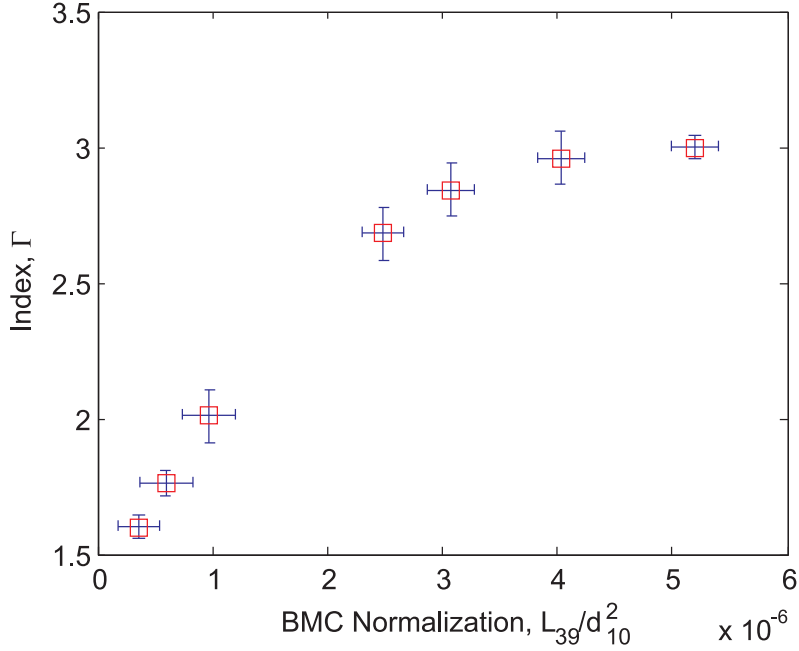}
\caption{Correlation of the photon index $\Gamma$ ($=\alpha+1$) %and the seed photon temperature (see the embedded panel) 
versus  the BMC normalization $N_{BMC}$ (proportional to mass accretion rate) in units of $L_{39}/D^2_{10}$. 
}
\label{saturation}
\end{figure}

\newpage
\begin{figure}[ptbptbptb]
\includegraphics[scale=1.0, angle=0]{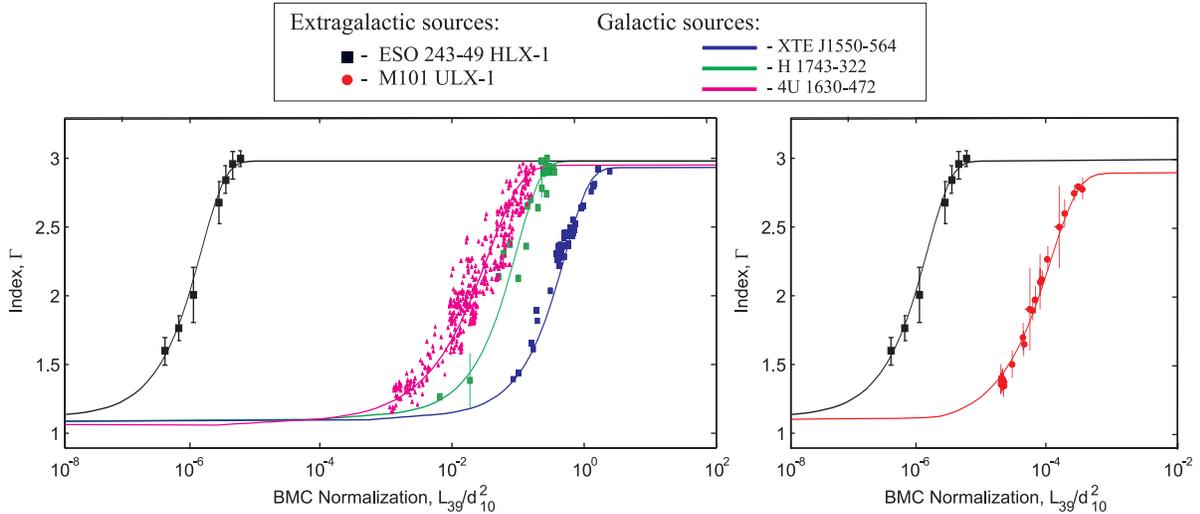}
\caption{Scaling of the photon index $\Gamma$ versus the normalization $N_{BMC}$ for ESO 243--49 HLX--1 (black 
points indicate the target source) using the correlations for the {Galactic} reference sources,  4U~1630-472, XTE J1550-564, and H1743-322 (pink, blue, and green, left panel) and comparison of $\Gamma - N_{BMC}$ correlations for extragalactic sources, ESO 243-49 HLX--1 and M101~ULX-1 (red  points, see right panel)
}
\label{three_scal}
\end{figure}

\end{document}